\documentclass[12pt]{article}
\usepackage{mathtext}
\usepackage[cp1251]{inputenc}
\usepackage{amsmath}
\usepackage{bm}
\usepackage[english,ukrainian]{babel}
\usepackage{color}
\usepackage{cite}
\begin{document}

\begin{center}
\begin{large}
{\bf Deformed Heisenberg algebras of different types with preserved weak equivalence principle}
\end{large}
\end{center}

\centerline {Kh. P. Gnatenko \footnote{E-Mail address: khrystyna.gnatenko@gmail.com}, V. M. Tkachuk \footnote{E-Mail address: voltkachuk@gmail.com}}
\medskip
\centerline {\small \it Ivan Franko National University of Lviv, Professor Ivan Vakarchuk Department for Theoretical Physics,}
\centerline {\small \it 12 Drahomanov St., Lviv, 79005, Ukraine}

\abstract{In the paper a review of results for recovering of the weak equivalence principle in a space with deformed commutation relations for operators of coordinates and momenta is presented. Different types of deformed algebras leading to a space quantization are considered among them noncommutative algebra of canonical type, algebra of Lie type, nonlinear deformed algebra with arbitrary function of deformation depending on momenta. A motion of a particle and a composite system in gravitational field is examined and the implementation of the weak equivalence principle is studied. The principle is preserved in quantized space if we consider parameters of deformed algebras to be dependent on mass.  It is also shown that dependencies of parameters of deformed algebras on mass lead to preserving of the properties of the kinetic energy in quantized spaces and solving of the problem of significant effect of space quantization on the motion of macroscopic bodies (the problem is known as the soccer-ball problem).


  PACS number(s): 02.40.Gh, 03.65.-w
}

\section{Introduction}

 Deformed commutation relations for coordinates and momenta were firsty proposed by Heisenberg. The author of the first paper with formalization of the idea of deformed commutation relations is Snyder \cite{Snyder}. It is worth noting that a grate interest for studies of different types of deformed algebras leading to the minimal length follows from investigations in the String Theory and Quantum Gravity (see, for instance,  \cite{Gross,Maggiore}).

 The Snyder's algebra is well known and studied (see, for example, \cite{Romero08,Mignemi11,Mignemi14,Lu,GnatenkoEPL19}).  The algebra in a nonrelativistic case reads
\begin{eqnarray}
[X_i,X_j]=i\hbar\beta(X_iP_j-X_jP_i),  \label{S1}\\{}
[X_i,P_j]=i\hbar(\delta_{ij}+\beta P_iP_j), \\{}
 [P_i,P_j]=0. \label{S2}
\end{eqnarray}

Also, well studied algebra leading to a minimal length is deformed algebra proposed by Kempf  (see, for instance, \cite{Kempf,Benczik02,Benczik05,Kempf95,Roy1,Roy2,GnatenkoIJMPD19})
 \begin{eqnarray}
 [X_i,X_j]=i\hbar \frac{(2\beta-\beta^{\prime})+(2\beta+\beta^{\prime})\beta P^2}{1+\beta P^2}(P_iX_j-P_jX_i),\label{K1}\\{}
[X_i,P_j]=i\hbar(\delta_{ij}(1+\beta P^2)+\beta^{\prime}P_iP_j),\\{}
[P_i,P_j]=0,\label{K2}
\end{eqnarray}
where $\beta$, $\beta^{\prime}$ are constants. In the space the minimal length is defined by the parameters of deformations and it reads $\hbar\sqrt{\beta+\beta^{\prime}}$.

It is worth noting that algebras  (\ref{S1})-(\ref{S2}),  (\ref{K1})-(\ref{K2}) are not invariant under translations in the configurational space.
Deformed algebra characterized by the following commutation relation
 \begin{eqnarray}
[X_i,X_j]=0,\label{C1}\\{}
[X_i,P_j]=i\hbar(\delta_{ij}(1+\beta P^2)+2\beta P_iP_j),\\{}
[P_i,P_j]=0,\label{C2}
\end{eqnarray}
describes uniform space.
This algebra can be obtained from (\ref{K1})-(\ref{K2}) up to the first order in the parameter of deformation, considering particular case $\beta^{\prime}=2\beta$.  We can also write deformed algebra
 \begin{eqnarray}
[X_i,X_j]=0,\label{9C1}\\{}
[X_i,P_j]=i\hbar\sqrt{1+\beta P^2}(\delta_{ij}+\beta P_iP_j),\\{}
[P_i,P_j]=0, \label{9C2}
\end{eqnarray}
which is invariant upon translations in the configurational space and leads to the minimal length (see \cite{Tk3}).

 In more general case one can consider the following commutation relations for coordinates and momenta
\begin{eqnarray}
[X_i,P_j]=i\hbar {F}_{ij}(\sqrt{\beta} P_1, \sqrt{\beta} P_2, \sqrt{\beta} P_3),\label{qggd}
\end{eqnarray}
 where ${F}_{ij}(\sqrt{\beta} P_1, \sqrt{\beta} P_2, \sqrt{\beta} P_3)$  are deformation functions. For preserving of the time-reversal symmetry and for invariance upon the parity transformations the functions have to be even
\begin{eqnarray}
F_{ij}(-\sqrt{\beta} P_1, -\sqrt{\beta} P_2, -\sqrt{\beta} P_3)=F_{ij}(\sqrt{\beta} P_1, \sqrt{\beta} P_2, \sqrt{\beta} P_3).
  \end{eqnarray}
In the literature algebra  (\ref{qggd}) with
\begin{eqnarray}
{F}_{ij}(\sqrt{\beta} P_1, \sqrt{\beta} P_2, \sqrt{\beta} P_3)=\delta_{ij}-\sqrt{\beta}\left(P\delta_{ij}+\frac{P_iP_j}{P}\right)+\nonumber\\+\beta(P^2\delta_{ij}+3P_iP_j),
 \end{eqnarray}
 was considered to describe a space with minimal length and maximal momentum \cite{Ali}. Also one-dimensional algebras
   \begin{eqnarray}
[X,P]=i\hbar F(\sqrt{\beta}|P|),\label{qgd}
\end{eqnarray}
were examined \cite{n,m}.  In (\ref{qgd})  $F(\sqrt{\beta}|P|)$ is a  deformation function, $\beta$ is a parameter, $\beta\geq0$, $F(0)=1$.

In the case of $F(\sqrt{\beta}|P|)=1+\beta P^2$, from deformed commutation relation (\ref{qgd}) follows  well known
generalized uncertainty principle (GUP)
\begin{equation}
\Delta X\geq\frac{\hbar}{2}\left(\frac{1}{\Delta P}+\beta\Delta P\right),\label{GUP}
\end{equation}
  leading to the minimal length $X_{min}=\hbar\sqrt{\beta}$.

  Also, other cases of the deformation functions leading to a minimal length and to a minimal momentum have been studied. Namely,  in \cite{Pedram12,Pedram122} the authors proposed to consider $F(\sqrt{\beta}|P|)=1/(1-\beta P^2)$. In the paper   \cite{Won18} it was chosen  $F(\sqrt{\beta}|P|)$ to be  $F(\sqrt{\beta}|P|)=(1-\sqrt{\beta}|P|)^2$. In \cite{Won19} the case of $F(\sqrt{\beta}|P|)=1/(1-\sqrt{\beta}|P|)$ was examined. The minimal length and the minimal momentum are defined by the parameter of deformation and are proportional to $\hbar \sqrt{\beta}$ and $1/\sqrt{\beta}$, respectively \cite{Ali,Pedram12,Won18,Won19}.

Algebras it which commutators for operators of coordinates and momenta are modified and give constants are known as noncommutative algebras of canonical type. In a general case these algebra reads
\begin{eqnarray}
[X_i,X_j]=i\hbar\theta_{ij},\label{33D1p}\\{}
[X_i,P_j]=i\hbar(\delta_{ij}+\sigma_{ij}),\label{sscan}\\{}
[P_i,P_j]=i\hbar \eta_{ij},\label{33D2p}
\end{eqnarray}
where $\theta_{ij}$ are parameters of coordinate noncommutativity, $\eta_{ij}$ are parameters of momentum noncommutativity and  $\sigma_{ij}$ are constants. Noncommutativity of coordinates leads to a minimal length. From the noncommutativity of momenta follows existence of the minimal momentum \cite{GnatUK}. Because of simplicity of the algebra it has received much attention  \cite{Djemai,
Alavi,Bertolami,Smailagic,Li,Acatrinei,Giri,Geloun,Bertolami1,Bastos}. More complicated type of noncommutative algebras are that of Lie type
  \begin{eqnarray}
 [X_{i},X_{j}]=i\hbar\theta^{k}_{ij}X_k,\label{0}
 \end{eqnarray}
 where $\theta^k_{ij}$ are constants  \cite{DaszkiewiczPRD,Lukierski,Lukierski18,Miao}.


An important problem is a construction of deformed algebra which leads to a space quantization and does not cause violation of fundamental physical laws and principles. For instance, well known problem in the frame of deformed algebras of different types is violation of the weak equivalence principle or Galilean equivalence principle or universality of free fall principle. Deformation of commutation relation for coordinates and momenta leads to dependence of velocity and position of a point mass in a gravitational field  on  mass. In the case of algebras with noncommutativity of coordinates of canonical type the equivalence principle was considered in \cite{Bastos1,Bertolami2,GnatenkoPLA13,Saha,GnatenkoPLA17,GnatenkoIJTP18}. More general case of noncommutativity of coordinates and noncommutativity of momenta was examined in \cite{Bastos1,Bertolami2,GnatenkoPLA17,GnatenkoIJTP18} and the problem of the ununiversality of free fall in the space was studied. In paper \cite{Bertolami2} it was concluded that the equivalence principle holds in the quantized space in the sense that an accelerated frame of reference is locally equivalent to
gravitational field, unless parameters of noncommutativity are anisotropic ($\eta_{xy}\neq$$\eta_{xz}$).
Generalized uncertainty relations preserving of the equivalence principle were studied in \cite{Lake_2019}.

In the paper we present the way to recover the weak equivalence principle in a spaces characterized by different types of deformed algebras, among them are
 noncommutative algebra of canonical type, noncommutative algebra of Lie type, the case of nonlinear deformed algebra with arbitrary deformation function depending on momentum. The solution of the problem on the basis of the idea of dependence of the parameters of deformed algebras on mass.
 It is important to stress the the idea  leads also to recovering of the properties of a kinetic energy and solving of well known soccer-ball problem (the problem of description of motion of a macroscopic body) in a space with minimal length.

The structure of the paper is the following. In Section \ref{G_sec2}  a  space with GUP is considered (\ref{qgd}), (\ref{qggd}) and the implementation of the weak equivalence principle in the case of nonlinear deformed algebras is recovered.  In Section \ref{G_sec3} a noncommutative algebra of canonical type is examined. The influence of noncommutativity of coordinates and noncommutativity of momenta on the E\"otv\"os parameter  for Sun-Earth-Moon system is found. Relations for the parameters of nocommutativity with mass for preserving of the weak equivalence principle are found.  Noncommutative algebra which is rotationally- and time-reversal invariant and does not lead to violation of the weak equivalence principle is studied  in Section \ref{G_sec5}. Implementation of the Galilean equivalence principle in a space with noncommutative algebra of Lie type is considered in Section \ref{G_sec6}.  
Section \ref{G_sec7} is devoted to conclusions.

\section{Preserving of the weak equivalence principle in a space with GUP}\label{G_sec2}

\subsection{Motion in gravitational field in a space with nonlinear deformed algebras}

As a first step of studying of the weak equivalence principle in spaces with nonlinear deformed algebras let us consider a one-dimensional case of algebra
with arbitrary function of deformation dependent on momenta (\ref{qgd}).
Relation (\ref{qgd}) corresponds to  the following deformed Poisson bracket
\begin{eqnarray}
\{X,P\}= F(\sqrt{\beta}|P|).\label{gd}
\end{eqnarray}

For a particle with mass $m$  in gravitational field $V(X)$, writing Hamiltonian
\begin{eqnarray}
H=\frac{P^2}{2m}+mV(X),\label{hhham}
\end{eqnarray}
and taking into account deformation of the Poisson brackets we find  equations of motion as
\begin{eqnarray}
\dot{X}=\{X,H\}=\frac{P}{m}{F}(\sqrt{\beta} |P|),\label{9eqm1}\\
\dot{P}=\{P,H\}=-m \frac{\partial V (X)}{\partial X}{F}(\sqrt{\beta} |P|).\label{9eqm2}
\end{eqnarray}
On the basis of the obtained expressions we can conclude that even if we consider in  (\ref{hhham})
the inertial mass to be equal to the gravitational mass the motion of a particle in gravitation field in a space with GUP depends on its mass and the weak equivalence principle is violated.

From equations (\ref{9eqm1}), (\ref{9eqm2}) follows that the motion of a particle in gravitational field in the space  (\ref{qgd}) depends on its mass. So, deformation of commutation relation (\ref{qgd}) leads to violation of the weak equivalence principle.

One face the same problem in three-dimensional case of deformed algebra (\ref{qggd})
and deformed Poisson brackets
\begin{eqnarray}
\{X_i,P_j\}= {F}_{ij}(\sqrt{\beta} P_1,\sqrt{\beta} P_2, \sqrt{\beta} P_3),\label{ggd}\\
\{X_i,X_j\}=\{P_i,P_j\}=0.\label{99ggd}
\end{eqnarray}
Here we would like to note that we consider the ordinary Poisson brackets $\{X_i,X_j\}$, and  $\{P_i,P_j\}$ (\ref{99ggd}) because in this case the deformed algebra  (\ref{ggd}), (\ref{99ggd}) is invariant with respect to translations in the configurational space.
Similarly as in the one-dimensional case we study a particle with Hamiltonian  $H=\sum_iP^2_i/{2m}+mV(\bf{X})$. Using (\ref{ggd}), (\ref{99ggd}) the equations of motion of the particle in the gravitational field read
\begin{eqnarray}
\dot{X}_i=\sum_i\frac{P_j}{m}{F}_{ij}(\sqrt{\beta} P_1, \sqrt{\beta} P_2, \sqrt{\beta} P_3),\label{eqm111}\\
\dot{P}_i= - m \sum_j\frac{\partial V (\bf{X})}{\partial X_j}{F}_{ij}(\sqrt{\beta} P_1, \sqrt{\beta} P_2, \sqrt{\beta} P_3).\label{eqm222}
\end{eqnarray}
On the basis of the obtained results we conclude that the weak equivalence principle is violated.

It is important to stress that the deformation of the commutation relations causes a grate  corrections to the E\"otv\"os parameter and grate violation of the weak equivalence principle.
 For instance,
in the case of uniform field
$V(X)=-gX$ ($g$ is the  gravitational acceleration)  equations of motions (\ref{9eqm1}), (\ref{9eqm2}) transform to
\begin{eqnarray}
\dot{X}=\frac{P}{m}{F}(\sqrt{\beta} |P|),\label{99eqm1}\\
\dot{P}=m g{F}(\sqrt{\beta} |P|),\label{99eqm2}
\end{eqnarray}
and expression for the acceleration written up to the first order in the parameter of deformation
is as follows
\begin{eqnarray}
\ddot{X}=g+3{F}^{\prime}(0)g\sqrt{\beta} m|\upsilon|+(2{F}^{\prime\prime}(0)-({F}^{\prime}(0))^2)g \beta m^2 \upsilon^2, \label{resaks}
\end{eqnarray}
  where $F^{\prime}(x)=d F/d x$, $F^{\prime\prime}(x)=d^2 F/d x^2$ and  $\upsilon$ is a  velocity of motion in gravitational field $V(X)=-gX$  in the case of $\beta=0$.
On the basis of (\ref{resaks}), for particles   with masses $m_1$, $m_2$ the E\"otv\"os parameter reads
\begin{eqnarray}
\frac{\Delta a}{a}=\frac{2(\ddot X^{(1)}-\ddot X^{(2)})}{\ddot X^{(1)}+\ddot X^{(2)}}=3{F}^{\prime}(0)|\upsilon|\sqrt{\beta} (m_1-m_2)+(2{F}^{\prime\prime}(0)-({F}^{\prime}(0))^2)\upsilon^2\beta (m^2_1-m^2_2).\label{eta}
\end{eqnarray}
If we consider the minimal length to be equal to the Planck length  $\hbar \sqrt{\beta}=l_P$ we obtain
 \begin{eqnarray}
\frac{\Delta a}{a}=3{F}^{\prime}(0)\frac{|\upsilon|}{c}\frac{(m_1-m_2)}{m_P}+(2{F}^{\prime\prime}(0)-({F}^{\prime}(0))^2)\frac{\upsilon^2}{c^2} \frac{(m^2_1-m^2_2)}{m_P^2}\label{gv},
\end{eqnarray}
where $c$ is the speed of light, $G$ is the gravitational constant, $m_P$ is the Planck mass \cite{GnatenkoMPLA20}.

For bodies with masses $m_1=1$ kg, $m_2=0.1$ kg  and $F(\sqrt{\beta} |P|)=1+\beta P^2$ the E\"otv\"os parameter posses large value $\Delta a/a\approx0.1$. Such a violation of the weak equivalence principle could be easily seen at an experiment. But we know that the equivalence principle holds with hight  precision, for instance from the Lunar Laser ranging experiment follows  $\Delta a/a=(-0.8\pm1.3)\cdot10^{-13}$ \cite{Williams}.

The problem is solved if parameter of deformation $\beta$  is considered to be dependent on mass
as follows
\begin{eqnarray}
\sqrt{\beta_a} m_a=\gamma=\textrm{const}. \label{ccc}
\end{eqnarray}
Here constant $\gamma$  which is the same for different particles is introduced \cite{GnatenkoMPLA20,Tk1,Tk2}.

If relation holds (\ref{ccc}) the E\"otv\"os parameter (\ref{eta}) is equal to zero and   equations (\ref{9eqm1}), (\ref{9eqm2}) transform to
\begin{eqnarray}
\dot{X}=\frac{P}{m}{F}\left(\gamma \frac{|P|}{m}\right),\label{eqm11}\\
\frac{\dot{P}}{m}= -\frac{\partial V (X)}{\partial X}{F}\left(\gamma \frac{|P|}{m}\right).\label{eqm22}
\end{eqnarray}
In (\ref{eqm11}), (\ref{eqm22})  we have that the mass is present  only in expression $P/m$. So,   $X(t)$, $P(t)/m$ do not depend on mass and the problem of violation of the weak equivalence principle is solved  \cite{GnatenkoMPLA20,Tk1}.

The same conclusion can be done in three-dimensional case. In the case of preserving of the condition (\ref{ccc}), introducing  ${P}_i^{\prime}={P}_i/m$ from (\ref{eqm111}), (\ref{eqm222}) we have
\begin{eqnarray}
\dot{X}_i=\sum_j{P^{\prime}_j}{F}_{ij}\left(\gamma P_{1}^{\prime}, \gamma P_{2}^{\prime}, \gamma P_{3}^{\prime}\right),\label{9eqm111c}\\
\dot{P}^{\prime}_i= -\sum_j\frac{\partial V (\bf{X})}{\partial X_j}{F}_{ij}\left(\gamma P_{1}^{\prime}, \gamma P_{2}^{\prime}, \gamma P_{3}^{\prime}\right).\label{9eqm222c}
\end{eqnarray}
So, the mass is canceled in  (\ref{9eqm111c}), (\ref{9eqm222c}) and  the motion in gravitational field does not depend on mass, the weak equivalence principle is preserved.

Let us recall that we considered deformed algebra with ordinary relations for $\{X_i,X_j\}$, $\{P_i,P_j\}$ (\ref{99ggd}).
But even in the case of more complicated deformed algebra the idea of dependence of  parameters of deformation on mass gives a possibility to recover the weak equivalence principle. For instance, in the case of the following commutation relations
\begin{eqnarray}
[X_i,X_j]=G(P^2)(X_iP_j-X_jP_i),\label{cgen1}\\{}
[X_i,P_j]=f(P^2)\delta_{ij}+F(P^2)P_iP_j,\label{cgen3}\\{}
[P_i,P_j]=0.\label{cgen2}
\end{eqnarray}
Algebra (\ref{cgen1})-(\ref{cgen2}) is generalization of well known Snyder (\ref{S1})-(\ref{S2}) and  Kempf (\ref{K1})-(\ref{K2}) algebras.
Functions $G(P^2)$, $F(P^2)$, $f(P^2)$  in (\ref{cgen1})-(\ref{cgen2}) cannot be chosen independently \cite{Frydryszak}.  From the  Jacobi identity  follows the following relation
\begin{eqnarray}
f(F-G)-2\frac{\partial f}{\partial P}(f+FP^2)=0,
\end{eqnarray}

Let us study the weak equivalence principle in quantized space  (\ref{cgen1})-(\ref{cgen2}).
Considering a particle in gravitational field with Hamiltonian  $H=\sum_i\frac{P^2_i}{2m}+mV(\bf{X})$  in a space with deformed algebra  (\ref{cgen1})-(\ref{cgen2}) and parameter of deformation satisfying condition (\ref{ccc}), we can write the equations of motion  as follows
\begin{eqnarray}
\dot{X}_i=P^{\prime}_i\tilde{f}(\gamma^2 (P^{\prime})^2)+\gamma^2\sum_j\frac{\partial V (\bf{X})}{\partial X_j}\tilde G(\gamma^2 (P^{\prime})^2)\left(X_i{P^{\prime}_j}-X_j{P^{\prime}_i}\right),\label{gen112}\\
{\dot{P}^{\prime}_i}= -\frac{\partial V (\bf{X})}{\partial X_i}\tilde{f}(\gamma^2(P^{\prime})^2)-\gamma^2\sum_j \frac{\partial V (\bf{X})}{\partial X_j}\tilde F(\gamma^2 (P^{\prime})^2){P^{\prime}_iP^{\prime}_j},\label{gen113}
\end{eqnarray}
where  $\tilde{f}(\beta P^2)$, $\tilde{F}(\beta P^2)$, $\tilde{G}(\beta P^2)$ are dimensionless functions corresponding to $f(P^2)$, $F(P^2)$, $G(P^2)$ respectively.
On the basis of equations (\ref{gen112}), (\ref{gen113}) we have that the weak equivalence principle is preserved in the general case of the deformed algebra  (\ref{cgen1})-(\ref{cgen2})  due to condition (\ref{ccc})  \cite{GnatenkoMPLA20}.

In the next subsection in addition we will show that  with the help of relation  (\ref{ccc})  the properties of the kinetic energy can be preserved in the frame of the deformed algebra.

\subsection{Properties of the kinetic energy in a space with GUP and dependence of parameter of deformation on mass}

 Using  relation of momenta with velocity (\ref{eqm11}), the kinetic energy of a free particle (a body) of mass $m$ in the space with GUP (\ref{gd}) up to the first order in  $\beta$ reads
 \begin{eqnarray}
T=\frac{P^2}{2m}=\frac{m \dot X^2}{2}-{F}^{\prime}(0){\sqrt{\beta} m^2 |\dot X|\dot X^2}+(5({F}^{\prime}(0))^2-{F}^{\prime\prime}(0))\frac{\beta m^3 \dot X^4}{2}.\label{T}
\end{eqnarray}
On the other hand from the additivity property for a system of $N$ particles with masses $m_a$ that move with the same velocities we can write \begin{eqnarray}
T=\sum_a T_a=\frac{m \dot X^2}{2}-{F}^{\prime}(0){\sqrt{\beta} |\dot X|\dot X^2}\sum_a m_a^2+(5({F}^{\prime}(0))^2-{F}^{\prime\prime}(0))\frac{\beta  \dot X^4}{2}\sum_a m_a^3,\label{sT}
\end{eqnarray}
where $m=\sum_am_a$
\begin{eqnarray}
T_a=\frac{m_a \dot X_a^2}{2}-{F}^{\prime}(0){\sqrt{\beta} m_a^2 |\dot X_a|\dot X_a^2}+(5({F}^{\prime}(0))^2-{F}^{\prime\prime}(0))\frac{\beta m_a^3 \dot X_a^4}{2},\label{Ta}
\end{eqnarray}
and we take into account  $\dot X_a=\dot X$.
The obtained results (\ref{sT}), (\ref{T}) are not the same. Note that $m^2=(\sum_a m_a)^2>\sum_a m_a^2$ and $m^3=(\sum_a m_a)^3>\sum_a m_a^3$. Therefore absolute values of the corrections to the kinetic energy (\ref{T}) of the first and the second order are  bigger than absolute values of the corrections in (\ref{sT}).

It is worth noting that for a system made of $N$ particles with the same masses we have
  \begin{eqnarray}
T=N\frac{m_a \dot X^2}{2}-N^2{F}^{\prime}(0){\sqrt{\beta} m_a^2 |\dot X|\dot X^2}+N^3(5({F}^{\prime}(0))^2-{F}^{\prime\prime}(0))\frac{\beta m_a^3 \dot X^4}{2},\label{Ti}\\
T=NT_a=N \frac{m_a \dot X^2}{2}-N\left({F}^{\prime}(0){\sqrt{\beta} m_a^2 |\dot X|\dot X^2} -(5({F}^{\prime}(0))^2-{F}^{\prime\prime}(0))\frac{\beta m_a^3 \dot X^4}{2}\right),\label{sTi}
\end{eqnarray}
here we take into account that  $m=Nm_a$.  The dependencies of corrections to the kinetic energy  on the number of particles $N$ caused by the deformation (\ref{gd}) are different in (\ref{Ti}) and (\ref{sTi}). Analyzing (\ref{Ti}) we have that corrections of the first and the second order in $\sqrt{\beta}$ are proportional to $N^2$ and $N^3$, respectively. Zero order term in (\ref{Ti}) is proportional to $N$. So, with increasing of the number of particles in a system (in a macroscopic body) corrections to the kinetic energy caused by GUP increase faster than the zero order term. From this follows that the effect of space quantization on the motion of a macroscopic body is significant \cite{GnatenkoMPLA20}.
The problem is similar to the problem of macroscopic bodies in Double Special Relativity
 which is well known as  the soccer-ball  problem \cite{Hossenfelder,Amelino-Camelia,Hossenfelder1}.

If condition (\ref{ccc})  is satisfied for parameters of deformation of particles $\sqrt{\beta_a} m_a=\gamma=\textrm{const}$ and for parameter of deformation of a composite system (macroscopic body) $\sqrt{\beta} m=\gamma=\textrm{const}$, the kinetic energy has additivity property, does not depend on composition besides it is proportional to the mass. On the basis of (\ref{Ti}) and (\ref{sTi}) we obtain
\begin{eqnarray}
T=
\frac{m \dot X^2}{2}-{F}^{\prime}(0){\gamma m |\dot X|\dot X^2}+(5({F}^{\prime}(0))^2-{F}^{\prime\prime}(0))\frac{\gamma^2 m \dot X^4}{2}.\label{rsT}
\end{eqnarray}
So, the problem of violation of the properties of the kinetic energy and the soccer-ball problem are solved due to relation (\ref{ccc}).

The same conclusion can be done in all orders in the parameter of deformation. If condition (\ref{ccc}) is satisfied we can rewrite (\ref{eqm11}) as
\begin{eqnarray}
\dot{X}=\frac{P}{m}{F}\left(\gamma \frac{|P|}{m}\right).\label{fvv}
\end{eqnarray}
From this equation we have that  $P/m$ is a function of velocity $\dot{X}$ and $\gamma$
\begin{eqnarray}
\frac{P}{m}=f(\dot{X},\gamma),\label{pp}
 \end{eqnarray}\
So,   $P$ is proportional to mass $m$. Using relation (\ref{pp}) we can rewrite the kinetic energy of the particle in the following form
 \begin{eqnarray}
T=\frac{P^2}{2m}=\frac{m (f(\dot{X},\gamma))^2}{2}.\label{tt}
\end{eqnarray}
Let us consider a system of $N$ particles which move with the same velocities. This is equivalent to the case of body divided to $N$ parts that can be considered as particles.
The kinetic energy of the system according to the additivity property can be written as
 \begin{eqnarray}
T=\sum_aT_a=\sum_a\frac{m_a (f(\dot{X},\gamma))^2}{2}=\frac{m (f(\dot{X},\gamma))^2}{2}.\label{add}
\end{eqnarray}
Here we use notation $m$ for the total mass of the system  $m=\sum_am_a$. Note that the same result (\ref{add}) we obtain on the basis of expression (\ref{tt}), substituting  $m=\sum_am_a$.
Another property of the kinetic energy, its independence of composition is also preserved due to relation (\ref{ccc}).
According to  (\ref{add})  the kinetic energy of a system is proportional to its total mass and does not depend on its composition as it is in the ordinary space (space with $\beta=0$).
So, besides recovering of the weak equivalence principle relation (\ref{ccc}) gives a possibility to preserve the properties of the kinetic energy in the space with GUP \cite{GnatenkoMPLA20,Tk1}.

Similarly in three-dimensional space  (\ref{ggd}) the kinetic energy has additivity property and is independent of composition if relation (\ref{ccc}) is satisfied.
For a free particle $H=\sum_{i}P_i^2/2m$ in the space (\ref{ggd}) we have
\begin{eqnarray}
\dot X_i=\{X,H\}=\sum_j\frac{P_j}{m}{F}_{ij}(\sqrt{\beta} P_1, \sqrt{\beta} P_2, \sqrt{\beta} P_3)=
\sum_j\frac{P_j}{m}{F}_{ij}\left(\gamma \frac{P_1}{m}, \gamma \frac{P_2}{m}, \gamma \frac{P_3}{m}\right).
\end{eqnarray}
Therefore if relation (\ref{ccc}) holds,  the values $P_i/m$  depend on velocities $\dot X_i$ and $\gamma$ and do not depend on mass
\begin{eqnarray}
\frac{P_i}{m}=f_i(\dot X_1, \dot X_2, \dot X_3, \gamma).
\end{eqnarray}
 So,  the kinetic energy of a particle with mass $m$ can be written as
\begin{eqnarray}
T=\sum_i\frac{m(f_i(\dot X_1, \dot X_2, \dot X_3, \gamma))^2}{2}.\label{ad0}
\end{eqnarray}
For a system of particles, which move with the same velocities according to the additivity property we can write
\begin{eqnarray}
T=\sum_a T_a=\sum_a\sum_i\frac{m_a(f_i(\dot X_1, \dot X_2, \dot X_3, \gamma))^2}{2}=\sum_i\frac{m(f_i(\dot X_1, \dot X_2, \dot X_3, \gamma))^2}{2},\label{ad}
\end{eqnarray}
here $m=\sum_am_a$. Result (\ref{ad}) corresponds to (\ref{ad0}).
 So, the properties of the kinetic energy are satisfied in all orders in the parameter of deformation if one considers dependence of the parameters of deformation corresponding to particles and macroscopic bodies on their masses (\ref{ccc}) \cite{GnatenkoMPLA20}.

According to condition (\ref{ccc}) parameters of deformation of macroscopic bodies are less then that corresponding to elementary particles. From (\ref{ccc}) the parameter of deformation of macroscopic body reads
\begin{eqnarray}
{\beta}={\beta_e} \frac{m^2_e}{m^2},\label{rfe}
\end{eqnarray}
where  $m_e$, $\beta_e$ are the mass and the parameter of deformation of an elementary particle.
 On the basis of (\ref{rfe}) we can conclude that there is a reduction  by the factor $m^2_e/m^2$  of the parameter of macroscopic body $\beta$  with respect to parameter of deformation $\beta_e$ corresponding to an elementary particle. Because of this reduction the problem of macroscopic bodies does not appear.

At the end of this section we would like to note that if relation (\ref{ccc}) is satisfied for the parameter of deformation of macroscopic body the motion of the body in gravitational field in a space with GUP does not depend on its mass and composition and the weak equivalence principle is satisfied.

\section{Motion in gravitational field in noncommutative phase space}\label{G_sec3}

\subsection{Recovering of the weak equivalence principle in a space with noncommutativity of coordinates and noncommutativity of momenta}
In a two-dimensional space with noncommutativity of coordinates and noncommutativity of momenta of canonical type the commutation relations for operators of coordinates and operators of momenta are as follows
 \begin{eqnarray}
[X_{1},X_{2}]=i\hbar\theta,\label{alI0}\\{}
[X_{i},P_{j}]=i\hbar\delta_{ij},\\{}
[P_{1},P_{2}]=i\hbar\eta,\label{alI1}
\end{eqnarray}
where $\theta$, $\eta$ are parameters of noncommutativity $i,j=(1,2)$.

 Let us consider an influence of noncommutativity of coordinates and noncommutativity of momenta on a motion of a particle in uniform gravitational field with Hamiltonian
 \begin{eqnarray}
 H=\frac{P_{1}^{2}}{2m}+\frac{P_{2}^{2}}{2m}-mgX_1,\label{hamun}
 \end{eqnarray}
  and examine the weak equivalence principle  \cite{GnatenkoPLA13,GnatenkoPLA17,GnatenkoIJTP18}.

Poisson brackets that correspond to relations of the deformed algebra  (\ref{alI0})-(\ref{alI1}) read
 \begin{eqnarray}
\{X_{1},X_{2}\}=\theta,\label{palI0}\\{}
\{X_{i},P_{j}\}=\delta_{ij},\\{}
\{P_{1},P_{2}\}=\eta.\label{palI1}
\end{eqnarray}
The definition of the deformed Poisson brackets is as follows
\begin{eqnarray}
\{f,g\}= \sum_i\left(\frac{\partial f}{\partial X_i}\frac{\partial g}{\partial P_i}-\frac{\partial f}{\partial P_i}\frac{\partial g}{\partial X_i}\right)+\theta\left(\frac{\partial f}{\partial X_1}\frac{\partial g}{\partial X_2}-\frac{\partial f}{\partial X_2}\frac{\partial g}{\partial X_1}\right)+\nonumber\\+\eta\left(\frac{\partial f}{\partial P_1}\frac{\partial g}{\partial P_2}-\frac{\partial f}{\partial P_2}\frac{\partial g}{\partial P_1}\right).\label{def1}
\end{eqnarray}

One obtains the following equations of motion and  and expressions for the trajectory of the particle in the gravitational field in noncommutative phase space of canonical type
  \begin{eqnarray}
\dot{X}_1=\{X_1,H\}=\frac{P_{1}}{m},\label{form0220}\\
\dot{X}_2=\{X_2,H\}=\frac{P_{2}}{m}+mg\theta,\\
\dot{P_{1}}=\{P_1,H\}=mg+\eta\frac{P_2}{m},\\
\dot{P_{2}}=\{P_2,H\}=-\eta\frac{P_1}{m},\label{form0221}\\
X_{1}(t)=\frac{m\upsilon_{01}}{\eta}\sin\frac{\eta}{m}t+\left(\frac{m^2g}{\eta^2}-\frac{m^2g\theta}{\eta}+\frac{m\upsilon_{02}}{\eta}\right)\left(1-\cos\frac{\eta}{m}t\right)+X_{01},\label{form0023}\\
X_{2}(t)=\left(\frac{m^2g}{\eta^2}-\frac{m^2g\theta}{\eta}+\frac{m\upsilon_{02}}{\eta}\right)\sin\frac{\eta}{m}t-\frac{m\upsilon_{01}}{\eta}\left(1-\cos\frac{\eta}{m}t\right)-\frac{mg}{\eta}t+mg\theta t+X_{02}.\label{form0024}
\end{eqnarray}
Here we use notations $X_{01}$, $X_{02}$, $\upsilon_{01}$, $\upsilon_{02}$ for the initial coordinates and velocities of the particle,  $X_1(0)=X_{01}$, $X_2(0)=X_{02}$, $\dot{X}_1(0)=\upsilon_{01}$, $\dot{X}_2(0)=\upsilon_{02}$.

From the obtained results we can conclude that the motion of a particle in gravitational field depends on its mass. So, in noncommutative phase space of canonical type we also face a problem of violation of the weak equivalence principle. It can be solved in the case when parameters of noncommutativity depend on mass as
 \begin{eqnarray}
\theta m=\gamma=\textrm{const},\label{ondN}\\
\frac{\eta}{m}=\alpha=\textrm{const},\label{ondN2}
\end{eqnarray}
where $\gamma$, $\alpha$  are constants that posses the same values for different particles \cite{GnatenkoPLA17}.  Using (\ref{ondN}), (\ref{ondN2}), (\ref{form0023}), (\ref{form0024}) we have that the mass is canceled in the expressions for the trajectory of a particle in gravitational field in noncommutative phase space
 \begin{eqnarray}
X_{1}(t)=\frac{\upsilon_{01}}{\alpha}\sin \alpha t+\left(\frac{g}{\alpha^2}-\frac{g\gamma}{\alpha}+\frac{\upsilon_{02}}{\alpha}\right)\left(1-\cos \alpha t\right)+X_{01},\label{form0025}\\
X_{2}(t)=\left(\frac{g}{\alpha^2}-\frac{g\gamma}{\alpha}+\frac{\upsilon_{02}}{\alpha}\right)\sin \alpha t-\frac{\upsilon_{01}}{\alpha}\left(1-\cos \alpha t\right)-\frac{g}{\alpha}t+\gamma g t+X_{02}.\label{form0026}
\end{eqnarray}
and the problem of violation of the weak equivalence principle is solved \cite{GnatenkoPLA17}.

Here it is worth to add that in the case of a space with noncommutativity of coordinates $\theta\neq0$, $\eta\rightarrow0$ on the basis of (\ref{form0023}), (\ref{form0024}) we have that the trajectory of particle in uniform field is not affected by noncommutativity $X_{1}(t)={gt^2}/{2}+\upsilon_{01}t+X_{01}$, $X_{2}(t)=\upsilon_{02}t+X_{02}$, but for the momenta we have the following expressions
$P_1=m\dot{X}_1$, $P_2=m(\dot{X}_2+mg\theta)$. Note, that the momentum $P_2$ is not proportional to mass.
It is worth also  mentioning that for $\eta\rightarrow0$ expressions (\ref{form0023}), (\ref{form0024})  transform to well known result for the trajectory of a particle in uniform gravitational field in the ordinary space, $X_{1}(t)={gt^2}/{2}+\upsilon_{01}t+X_{01}$, $X_{2}(t)=\upsilon_{02}t+X_{02}$. At the same time, if relation (\ref{ondN}) is satisfied the proportionality of momentum to mass is recovered $P_2=m(\dot{X}_2+\gamma g)$.

Let us consider more general case. Let us study the motion of a composite system in nonuniform gravitational field in noncommutative phase space and examine the weak equivalence principle. For this purpose we need to generalize  relations of noncommutative algebra for coordinates and momenta for different particles. We have
\begin{eqnarray}
\{X_{1}^{(a)},X_{2}^{(b)}\}=\delta^{ab}\theta_{a},\label{pal0}\\{}
\{X_{i}^{(a)},P_{j}^{(b)}\}=\delta^{ab}\delta_{ij},\\{}
\{P_{1}^{(a)},P_{2}^{(b)}\}=\delta^{ab}\eta_{a},\label{pal1}
\end{eqnarray}
 where indices $a$, $b$ label the particles, $X_{i}^{(a)}$, $P_{i}^{(a)}$ are coordinates and momenta of the particle with index $a$, $i=(1,2)$, $j=(1,2)$. In (\ref{pal0})-(\ref{pal1}) we consider a general case when coordinates and momenta of different particles satisfy noncommutative algebra with different parameters of noncommutativity. We use notations  $\theta_{a}$, $\eta_{a}$ for  parameters of noncommutativity corresponding to a particle with index $a$. Also, in (\ref{pal0})-(\ref{pal1})  we assume that Poisson brackets for coordinates and momenta corresponding to different particles are equal to zero.

Let us consider a composite system made of $N$ particles with masses $m_a$.
 Defining the coordinates and momenta of the center-of-mass, coordinates and momenta of the relative motion as in the ordinary space
\begin{eqnarray}
\tilde{{\bf P}}=\sum_{a}{\bf P}^{(a)}, \ \ \tilde{{\bf X}}=\sum_{a}\mu_{a}{\bf X}^{(a)}, \label{905}\\
\Delta{\bf P}^{{a}}={\bf P}^{(a)}-\mu_{a}\tilde{{\bf P}},\ \
{\Delta\bf X}^{(a)}={\bf X}^{(a)}-\tilde{{\bf X}},\label{906}
\end{eqnarray}
(here ${\bf X}^{(a)}=(X^{(a)}_1,X^{(a)}_2)$, ${\bf P}^{(a)}=(P^{(a)}_1,P^{(a)}_2)$) and using (\ref{pal0})-(\ref{pal1}), one obtains the following relations
\begin{eqnarray}
\{\tilde{X}_1,\tilde{X}_2\}=\tilde{\theta}, \ \  \{\tilde{P}_1,\tilde{P}_2\}=\tilde{\eta},\label{9907}\\{}
\{\tilde{X}_i,\tilde{P}_j\}=\{\Delta{X}_i,\Delta{P}_j\}=\delta_{ij},\label{9908}{}\\
\{\Delta{X}_1^{(a)},\Delta{X}_2^{(b)}\}=-\{\Delta{X}_2^{(a)},\Delta{X}_1^{(b)}\}=
=\delta^{ab}\theta_{a}-\mu_{a}\theta_{a}-\mu_{b}\theta_{b}+\tilde{\theta},{}\\
\{\Delta{P}_1^{(a)},\Delta{P}_2^{(b)}\}=-\{\Delta{P}_2^{(a)},\Delta{P}_1^{(b)}\}=
=\delta^{ab}\eta_a-\mu_b\eta_a-\mu_a\eta_b+\mu_a\mu_b\tilde{\eta}.{}
\end{eqnarray}
Parameters $\tilde{\theta}$, $\tilde{\eta}$ are defined as
\begin{eqnarray}
\tilde{\theta}=\frac{\sum_{a}m_{a}^{2}\theta_{a}}{(\sum_{b}m_{b})^{2}},\label{9eff}\\
\tilde{\eta}=\sum_{a}\eta_a,\label{9eff2}
\end{eqnarray}
 and are called as effective parameters of noncommutativity. So, coordinates and momenta of the center-of-mass of composite system satisfy noncommutative algebra with effective parameters which depend on the masses of particles forming it and on parameters of noncommutativity ${\theta}_a$, ${\eta}_a$ \cite{GnatenkoPLA17}.
It is important that the motion of the center-of-mass is not independent of the relative motion because of relations
 \begin{eqnarray}
 \{\tilde{X}_{1},\Delta X_{2}^{(a)}\}=-\{\tilde{X}_{2},\Delta X_{1}^{(a)}\}=\mu_{a}\theta_{a}-\tilde{\theta},\\{}
 \{\tilde{P}_1,\Delta{P}^{a}_2\}=-\{\tilde{P}_2,\Delta{P}^{a}_1\}=\eta_a-\mu_a\sum_{b}\eta_b.
\end{eqnarray}

The situation is changed if we consider conditions on the parameters of noncommutativity (\ref{ondN}), (\ref{ondN2}). In this case
 \begin{eqnarray}
 \{\tilde{X}_{1},\Delta X_{2}^{(a)}\}=-\{\tilde{X}_{2},\Delta X_{1}^{(a)}\}=0,\label{pzz}\\{}
 \{\tilde{P}_1,\Delta{P}^{a}_2\}=-\{\tilde{P}_2,\Delta{P}^{a}_1\}=0,\label{pzz1}
\end{eqnarray}
and we have that the motion of the center-of-mass is independent of the relative motion. Also due to relations (\ref{ondN}), (\ref{ondN2}) the effective parameters of  noncommutativity does not depend on the masses of particles forming the system and its composition. Using (\ref{9eff}), (\ref{9eff2}) and considering conditions (\ref{ondN}), (\ref{ondN2}) we obtain that the effective parameter of coordinate noncommutativity is proportional inversely to the total mass of the system \begin{eqnarray}
\tilde{\theta}=\frac{\gamma}{M}.
\end{eqnarray}
The effective parameter of momentum noncommutativity is proportional to the total mass of the system
\begin{eqnarray}
\tilde{\eta}=\alpha M.
\end{eqnarray}
So, conditions (\ref{ondN}), (\ref{ondN2}) are also satisfied for effective parameters of noncommutativity \cite{GnatenkoPLA17}.

 Let us examine the motion of a composite system in gravitational field in noncommutative phase space of canonical type taking into account the obtained results and conclusions about features of noncommutative algebra for coordinates and momenta of the center-of-mass and relative motion. We study the following Hamiltonian
 \begin{eqnarray}
 H=\frac{\tilde{{\bf P}}^{2}}{2M}+M V(\tilde{X}_1,\tilde{X}_2)+H_{rel}.
 \end{eqnarray}
Coordinates and momenta of the center-of-mass $\tilde{X}_i$, $\tilde{P}_i$  (\ref{905}), satisfy noncommutative algebra (\ref{9907}), (\ref{9908}) with  parameters $\tilde{\theta}$, $\tilde{\eta}$ given by (\ref{9eff}), (\ref{9eff2}), $M$  is the total mass of the system, the term $H_{rel}$ corresponds to the relative motion.

 If parameters of noncommutativity are related with mass  (\ref{ondN}), (\ref{ondN2}), the Poisson brackets for coordinates and momenta of the center-of-mass and relative motion are equal to zero (\ref{pzz}), (\ref{pzz1}), therefore
 \begin{eqnarray}
 \left\{\frac{\tilde{{\bf P}}^{2}}{2M}+M V(\tilde{X}_1,\tilde{X}_2),H_{rel}\right\}=0.
 \end{eqnarray}
 So,   equations of motion for the center-of-mass of a composite system in gravitational field read
 \begin{eqnarray}
\dot{\tilde{X}}_1=\frac{P_{1}}{M}+M\tilde{\theta}\frac{\partial V(\tilde{X}_1,\tilde{X}_2)}{\partial \tilde{X}_2}=
\frac{P_{1}}{M}+\gamma\frac{\partial V(\tilde{X}_1,\tilde{X}_2)}{\partial \tilde{X}_2},\label{form020}\\
\dot{\tilde{X}}_2=\frac{P_{2}}{M}-M\tilde{\theta}\frac{\partial V(\tilde{X}_1,\tilde{X}_2)}{\partial \tilde{X}_1}=\frac{P_{2}}{M}-\gamma\frac{\partial V(\tilde{X}_1,\tilde{X}_2)}{\partial \tilde{X}_1},\\
\dot{\tilde{P}}_{1}=-M\frac{\partial V(\tilde{X}_1,\tilde{X}_2)}{\partial \tilde{X}_1}+\tilde{\eta}\frac{P_2}{M}=
-M\frac{\partial V(\tilde{X}_1,\tilde{X}_2)}{\partial \tilde{X}_1}+\alpha P_2,\\
\dot{\tilde{P}}_{2}=-M\frac{\partial V(\tilde{X}_1,\tilde{X}_2)}{\partial \tilde{X}_2}-\tilde{\eta}\frac{P_1}{M}=-M\frac{\partial V(\tilde{X}_1,\tilde{X}_2)}{\partial \tilde{X}_2}-\alpha P_1.\label{form021}
\end{eqnarray}
Note that if conditions (\ref{ondN}), (\ref{ondN2}) are not satisfied the equations of motion of a composite system in gravitational field depend on  effective parameters of noncommutativity  (\ref{9eff}), (\ref{9eff2}) that are determined by the masses and parameters of noncommutativity of particles forming the system and depend on its composition. This causes to the violation of the weak equivalence principle.
If relations (\ref{ondN}), (\ref{ondN2})  are preserved the weak equivalence principle holds,
the motion of a composite system (a body) in gravitational field depends on the constants $\gamma$, $\alpha$ and does not depend on its mass and composition \cite{GnatenkoPLA17}.

Also, due to conditions (\ref{ondN}), (\ref{ondN2}) the properties of the kinetic energy are preserved in noncommutative phase space of canonical type. This will be shown in the next Subsection.

\subsection{Motion of composite system in gravitational filed and the properties of the kinetic energy}
Let us consider a composite system which is made of $N$ particles that move with the same velocities.
On the basis of (\ref{form0220})-(\ref{form0221}), considering the case when influence of relative motion on the
motion of the center-of-mass of the system is small,  for the composite system in uniform gravitational field we can write
\begin{eqnarray}
\tilde{P}_1=M\tilde{\upsilon}_{01}\cos\frac{\tilde{\eta}}{M}t+(M\tilde{\upsilon}_{02}+\frac{M^2g}{\tilde{\eta}}-M^2g\tilde{\theta})\sin\frac{\tilde{\eta}}{M}t,\label{form028}\\
\tilde{P}_2=-M\tilde{\upsilon}_{01}\sin\frac{\tilde{\eta}}{M}t+(M\tilde{\upsilon}_{02}+\frac{M^2g}{\tilde{\eta}}-M^2g\tilde{\theta})\cos\frac{\tilde{\eta}}{M}t-\frac{M^2g}{\tilde{\eta}},\label{form029}
\end{eqnarray}
where $M$ is the total mass of the system, $\tilde{\theta}$, $\tilde{\eta}$ are  effective parameters of noncommutativity corresponding to the system (\ref{9eff}), (\ref{9eff2}), $\tilde{\upsilon}_{01}$, $\tilde{\upsilon}_{02}$ are initial velocities of the center-of-mass of the system, $\dot{\tilde{X}}_1(0)=\dot{\tilde{\upsilon}}_{01}$, $\tilde{X}_2(0)=\tilde{\upsilon}_{02}$.
Using (\ref{form028})-(\ref{form029}) the kinetic energy of the system can be written in the following form
\begin{eqnarray}
T=\frac{\tilde{P}_1^2}{2M}+\frac{\tilde{P}_2^2}{2M}=
T_0+g^2M^3\left(\frac{1}{\tilde{\eta}^2}+\frac{\tilde{\theta}^2}{2}-\frac{\tilde{\theta}}{\tilde{\eta}}\right)+M^2g\tilde{\upsilon}_{02}\left(\frac{1}{\tilde{\eta}}-\tilde{\theta}\right)+\nonumber\\+
\frac{M^2g}{\tilde{\eta}}\left(\tilde{\upsilon}_{01}\sin\frac{\tilde{\eta}}{M}t+\left(\frac{M g}{\tilde{\eta}}-M g \tilde{\theta}+\tilde{\upsilon}_{02}\right)\cos\frac{\tilde{\eta}}{M}t\right).\label{form040}
\end{eqnarray}
According to the additivity property, taking into account that the velocities of particles are the same, we can write
\begin{eqnarray}
T=\sum_{a}T_{a}=\sum_{a}\frac{(P_1^{(a)})^2}{2m_a}+\frac{(P_2^{(a)})^2}{2m_a}=\nonumber\\
=\sum_{a}\left[T_{0a}+g^2m_a^3\left(\frac{1}{\eta_a^2}+\frac{\theta_a^2}{2}-\frac{\theta_a}{\eta_a}\right)+m_a^2g\tilde{\upsilon}_{02}\left(\frac{1}{\eta_a}-\theta_a\right)+\right.\nonumber\\\left.+
\frac{m_a^2g}{\eta_a}\left(\tilde{\upsilon}_{01}\sin\frac{\eta_a}{m_a}t+\left(\frac{m_a g}{\eta_a}-m_a g \theta_a+\tilde{\upsilon}_{02}\right)\cos\frac{\eta_a}{m_a}t\right)\right].\label{form041}
\end{eqnarray}
Expression (\ref{form041}) does not correspond to  (\ref{form040}). The properties of the kinetic energy are violated in noncommutative phase space. Namely, if parameters of noncommutativity are considered to be the same for different particles, one faces a problem of nonadditivity of the kinetic energy and its dependence on composition.
 Considering conditions (\ref{ondN}), (\ref{ondN2}) we can rewrite (\ref{form040}), (\ref{form041})  as
\begin{eqnarray}
T=
T_0+\sum_{a}m_{a}\left[g^2\left(\frac{1}{\alpha^2}+\frac{\gamma^2}{2}-\frac{\gamma}{\alpha}\right)+g\tilde{\upsilon}_{02}\left(\frac{1}{\alpha}-\gamma\right)+\right.\nonumber\\\left.+
\frac{ g}{\alpha}\left(\tilde{\upsilon}_{01}\sin \alpha t+\left(\frac{g}{\alpha}-g \gamma+\tilde{\upsilon}_{02}\right)\cos \alpha t\right)\right].\label{form044}
\end{eqnarray}
 On the basis of (\ref{form044}) we can conclude that the additivity property of the kinetic energy is preserved and the kinetic energy of a composite system does not depend on its composition \cite{GnatenkoPLA17}.

So, besides preserving of the weak equivalence principle in noncommutative phase space of canonical type conditions (\ref{ondN}), (\ref{ondN2}) give possibility to recover the properties of the kinetic energy, to consider the motion of the center-of-mass independently of the relative motion \cite{GnatenkoPLA13,GnatenkoPLA17,GnatenkoMPLA18}.

In the next Subsection using the obtained results we study effect of noncommutativity of coordinates and noncommutativity of momenta on the weak equivalence principle considering Sun-Earth-Moon system.

\subsection{Effect of noncommutativity on the E\"otv\"os parameter}

According to the Lunar laser ranging experiment the weak equivalence principle holds with accuracy
\begin{eqnarray}
\frac{\Delta a}{a}=\frac{2(a_E-a_M)}{a_E+a_M}=(-0.8\pm 1.3)\cdot10^{-13},\label{d34}
\end{eqnarray}
 (see \cite{Williams}). In  (\ref{d34}) $a_E$, $a_M$ are the free fall accelerations of Earth and the Moon toward the Sun when the bodies are at the same distance from the source of gravity.
On the basis of this result one can examine  conditions for the parameters of coordinates and momentum noncommutativity (\ref{ondN}), (\ref{ondN2})  proposed for preserving of the weak equivalence principle.
For this purpose we study influence of noncommutativity of coordinates and noncommutativity of momenta on the motion of the Earth and the Moon in the gravitational field  of the Sun.

We consider the following Hamiltonian
\begin{eqnarray}
 H=\frac{({\bf P}^E)^{2}}{2m_E}+\frac{({\bf P}^M)^{2}}{2m_M}-G\frac{m_E m_S}{R_{ES}}-G\frac{m_M m_S}{R_{MS}}-G\frac{m_M m_E}{R_{EM}}.
 \label{form12.5}
 \end{eqnarray}
The distances between bodies $R_{ES}$, $R_{MS}$, $R_{EM}$ in the case when the Sun is considered to be at the origin of the coordinate system  read
\begin{eqnarray}
R_{ES}=\sqrt{(X^E_1)^2+(X^E_2)^2}, \ \ R_{MS}=\sqrt{(X^M_1)^2+(X^M_2)^2}, \label{xxe}\\
R_{EM}=\sqrt{(X^E_1-X^M_1)^2+(X^E_2-X^M_2)^2}.\label{xxe}
 \end{eqnarray}
Coordinates and momenta $X^E_i$, $X^M_i$, $P^E_i$, $P^M_i$ correspond to the Earth and the Moon, $G$ is the gravitational constant,  $m_S$, $m_E$, $m_M$  are the masses of the Sun, the Earth and the Moon, respectively. It is worth noting that in (\ref{form12.5}) we consider the case when the inertial mass of the Earth (mass in the first term) is equal to its gravitational mass  (mass in the third and the fifth terms), also the inertial mass of the Moon (mass in the second term) is equal to its gravitational mass (mass in the fourth and the fifth terms).

In noncommutative phase space of canonical type we have the following Poisson brackets
 \begin{eqnarray}
 \{X_1^E,X_2^E\}=\theta_E, \ \ \{P^E_1,P^E_2\}=\eta_E,\ \ \{X^E_i,P^E_j\}=\delta_{ij},\label{dp}\\
 \{X_1^M,X_2^M\}=\theta_M, \ \ \{P^M_1,P^M_2\}=\eta_M,\ \ \{X^M_i,P^M_j\}=\delta_{ij},\\
 \{X_i^M,X_j^E\}= \{P_i^M,P_j^E\}=0,  \label{dp1}
 \end{eqnarray}
$\theta_E$, $\theta_M$, $\eta_E$, $\eta_M$ are parameters of coordinates and momentum noncommutativity corresponding to the Earth and the Moon.
Taking this into account we can write equations of motion \cite{GnatenkoIJTP18}
\begin{eqnarray}
\dot{X}^E_1=\frac{P^E_{1}}{m_E}+\theta_E\frac{Gm_Em_S X^E_2}{R_{ES}^{3}}+\theta_E\frac{Gm_Em_M (X^E_2-X^M_2)}{R_{EM}^{3}},\label{form020}\\
\dot{X}^E_2=\frac{P^E_{2}}{m_E}-\theta_E\frac{Gm_Em_S X^E_1}{R_{ES}^{3}}-\theta_E\frac{Gm_Em_M (X^E_1-X^M_1)}{R_{EM}^{3}},\label{form021}\\
\dot{P}^E_1=\eta_E\frac{P^E_{2}}{m_E}-\frac{Gm_Em_S X^E_1}{R_{ES}^{3}}-\frac{Gm_Em_M (X^E_1-X^M_1)}{R_{EM}^{3}},\label{form022}
\end{eqnarray}
\begin{eqnarray}
\dot{P}^E_2=-\eta_E\frac{P^E_{1}}{m_E}-\frac{Gm_Em_S X^E_2}{R_{ES}^{3}}-\frac{Gm_Em_M (X^E_2-X^M_2)}{R_{EM}^{3}},\label{form022}
\end{eqnarray}
\begin{eqnarray}
\dot{X}^M_1=\frac{P^M_{1}}{m_M}+\theta_M\frac{Gm_M m_S X^M_2}{R_{MS}^{3}}-\theta_M\frac{G m_E m_M (X^E_2-X^M_2)}{R_{EM}^{3}},\label{form023}
\end{eqnarray}
\begin{eqnarray}
\dot{X}^M_2=\frac{P^M_{2}}{m_M}-\theta_M\frac{Gm_M m_S X^E_1}{R_{MS}^{3}}+\theta_M\frac{G m_E m_M (X^E_1-X^M_1)}{R_{EM}^{3}},\label{form024}
\end{eqnarray}
\begin{eqnarray}
\dot{P}^M_1=\eta_M\frac{P^M_{2}}{m_M}-\frac{Gm_M m_S X^M_1}{R_{MS}^{3}}+\frac{Gm_Em_M (X^E_1-X^M_1)}{R_{EM}^{3}},\label{form025}
\end{eqnarray}
\begin{eqnarray}
\dot{P}^M_2=-\eta_M\frac{P^M_{1}}{m_M}-\frac{Gm_M m_S X^M_2}{R_{MS}^{3}}+\frac{Gm_Em_M (X^E_2-X^M_2)}{R_{EM}^{3}}.\label{form026}
\end{eqnarray}
On the basis of these equations accelerations of the Earth and the Moon can be found. Up to the first order in the parameters of coordinate and momentum noncommutativity we obtain
\begin{eqnarray}
\ddot{X}^E_1=-\frac{Gm_S X^E_1}{{ R}_{ES}^{3}}-\frac{Gm_M (X^E_1-X^M_1)}{R_{EM}^{3}}+\eta_E\frac{\dot{X}^E_{2}}{m_E} +{\theta_E}\frac{Gm_{S}m_E \dot{X}^E_{2}}{R_{ES}^{3}}+\nonumber
\end{eqnarray}
\begin{eqnarray}
+{\theta_E}\frac{Gm_{M}m_E }{R_{EM}^{3}}(\dot{X}^E_{2}-\dot{X}^M_{2})
-\theta_E\frac{3Gm_S m_E}{R_{ES}^5}({\bf R}_{ES}\cdot{\bf \dot{R}}_{ES}){X^E_2}-\nonumber\\-\theta_E\frac{3Gm_M m_E}{R_{EM}^5}({\bf R}_{EM}\cdot{\bf \dot{R}}_{EM})(X^E_2-X^M_2),\nonumber\\ \label{ddot1E}\\
\ddot{X}^M_1=-\frac{Gm_S X^M_1}{{ R}_{MS}^{3}}+\frac{Gm_E (X^E_1-X^M_1)}{R_{EM}^{3}}+\eta_M\frac{\dot{X}^M_{2}}{m_M} +{\theta_M}\frac{Gm_{S}m_M \dot{X}^M_{2}}{R_{MS}^{3}}-\nonumber\\-{\theta_M}\frac{Gm_{M}m_E }{R_{EM}^{3}}(\dot{X}^E_{2}-\dot{X}^M_{2})
-\theta_M\frac{3Gm_S m_M}{R_{MS}^5}({\bf R}_{MS}\cdot{\bf \dot{R}}_{MS}){X^M_2}+\nonumber\\+\theta_M\frac{3Gm_M m_E}{R_{EM}^5}({\bf R}_{EM}\cdot{\bf \dot{R}}_{EM})(X^E_2-X^M_2),\nonumber\\ \label{ddot1M}
\end{eqnarray}
where  ${\bf R}_{ES}(X_1^E,X_2^E)$, ${\bf R}_{MS}(X_1^M,X_2^M)$, ${\bf R}_{EM}(X_1^E-X_1^M,X_2^E-X_2^M)$ \cite{GnatenkoIJTP18}.

 In the case when the distance from the bodies to the Sun is the same, we can write $R_{MS}=R_{ES}=R$.
 For convenience we consider  $X_1$ axis to passe through the middle of ${\bf R}_{EM}$ and to be perpendicular to  ${\bf R}_{EM}$, $X_2$ axis to be  parallel to the ${\bf R}_{EM}$. Let us remind that we have chosen the origin of the frame of references at the Sun's center.
So, taking into account that $R_{EM}/R\sim10^{-3}$,  one obtains
\begin{eqnarray}
X^E_1=X^M_1=R\sqrt{1-\frac{R^2_{EM}}{4R^2}}\simeq R,\ \
X^E_2=-X^M_2=\frac{R_{EM}}{2}.
\end{eqnarray}
Note that
\begin{eqnarray}
\dot{X}^E_1=0, \ \ \dot{X}^M_1=\upsilon_M,\ \
\dot{X}^E_{2}=\dot{X}^M_{2}=\upsilon_E,
\end{eqnarray}
where $\upsilon_M$, $\upsilon_E$ are the orbital velocities of the Moon and the Earth.  So,  the free fall accelerations of the Moon and the Earth toward the Sun in the case when the bodies are at the same distance to it read
\begin{eqnarray}
a_E=\ddot{X}^E_1=-\frac{Gm_S}{R^{2}}+\eta_E\frac{\upsilon_E}{m_E} +{\theta_E}\frac{Gm_{S}m_E \upsilon_E}{R^{3}}\left(1-\frac{3R_{EM}}{2\upsilon_E R^2}({\bf R}_{ES}\cdot{\bf \dot{R}}_{ES})\right),\label{ddot1E}
\end{eqnarray}
\begin{eqnarray}
a_M=\ddot{X}^M_1=-\frac{Gm_S}{R^{2}}+\eta_M\frac{\upsilon_E}{m_M} +{\theta_M}\frac{Gm_{S}m_M \upsilon_E}{R^{3}}\left(1+\frac{3R_{EM}}{2\upsilon_E R^2}({\bf R}_{MS}\cdot{\bf \dot{R}}_{MS})\right).\label{ddot1M}
\end{eqnarray}

We have $R_{EM}/R\sim10^{-3}$, $\upsilon_M/\upsilon_E\sim10^{-2}$, therefore
\begin{eqnarray}
\frac{3R_{EM}({\bf R}_{ES}\cdot{\bf \dot{R}}_{ES})}{2\upsilon_E R^2}\sim10^{-6},\ \
 \frac{3R_{EM}({\bf R}_{MS}\cdot{\bf \dot{R}}_{MS})}{2\upsilon_E R^2}\sim10^{-5},
  \end{eqnarray}
  and the  last terms in the expressions for the accelerations (\ref{ddot1E}), (\ref{ddot1M}) can be neglected.  So,  for the E\"otv\"os parameter for the Earth and the Moon in noncommutative phase space we obtain the following result
\begin{eqnarray}
\frac{\Delta a}{a}=\frac{\upsilon_ER^2}{Gm_S}\left(\frac{\eta_E}{m_E}-\frac{\eta_M}{m_M}\right)+\frac{\upsilon_E}{R}\left(\theta_E m_E-\theta_M m_M\right)=\frac{\Delta a^{\eta}}{a}+\frac{\Delta a^{\theta}}{a}, \label{e}
\end{eqnarray}
where ${\Delta a^{\eta}}/{a}$, ${\Delta a^{\theta}}/{a}$ are corrections to the E\"otv\"os parameter caused by coordinate noncommutativity and momentum noncommutativity
\begin{eqnarray}
\frac{\Delta a^{\eta}}{a}=\frac{\upsilon_ER^2}{Gm_S}\left(\frac{\eta_E}{m_E}-\frac{\eta_M}{m_M}\right),\label{de}\\
\frac{\Delta a^{\theta}}{a}=\frac{\upsilon_E}{R}\left(\theta_E m_E-\theta_M m_M\right),\label{dt}
\end{eqnarray}
respectively.

It is important to stress that even if we consider the inertial masses of the bodies to be equal to their gravitational masses (see (\ref{form12.5})) the E\"otv\"os parameter is not equal to zero. Noncommutativity of coordinates and noncommutativity of momenta causes the violation of the weak equivalence principle. In addition it is worth to emphasize that parameters $\theta_E$, $\eta_E$,  $\theta_M$, $\eta_M$ correspond to macroscopic bodies, they are effective parameters of noncommutativity which depend on composition of the bodies and are defined as (\ref{9eff}), (\ref{9eff2}).
So, even for two bodies with the same masses but different compositions the E\"otv\"os-parameter is not equal to zero \cite{GnatenkoIJTP18}.

Let us introduce constants
\begin{eqnarray}
\alpha_E=\frac{\eta_E}{m_E},\ \ \alpha_M=\frac{\eta_M}{m_M},\ \
\gamma_E=\theta_E m_E,\ \ \gamma_M=\theta_M m_M,
\end{eqnarray}
and estimate the values $|\alpha_E-\alpha_M|$, $|\gamma_E-\gamma_M|$ on the basis of the Lunar laser ranging experiment results \cite{Williams}. We assume that the following inequality is satisfied
\begin{eqnarray}
\left|\frac{\Delta a^{\theta}+\Delta a^{\eta}}{a}\right|\leq2.1\cdot10^{-13}.
\end{eqnarray}
Here $2.1\cdot10^{-13}$  is the largest value  in (\ref{d34}) \cite{Williams}.
To estimate the orders of the values $|\alpha_E-\alpha_M|$, $|\gamma_E-\gamma_M|$ we consider  inequalities
\begin{eqnarray}
\left|\frac{\Delta a^{\theta}}{a}\right|\leq10^{-13},\ \
\left|\frac{\Delta a^{\eta}}{a}\right|\leq10^{-13}.\label{nr1}
\end{eqnarray}
From the inequalities, using (\ref{de}), (\ref{dt}), we find \cite{GnatenkoIJTP18}
\begin{eqnarray}
|\alpha_E-\alpha_M| \leq10^{-20}\textrm{s}^{-1},\ \
|\gamma_E-\gamma_M|\leq10^{-7}\textrm{s}.\label{uupte}
\end{eqnarray}

It is important to stress that considering conditions on the parameters of noncommutativity proposed in the previous section, namely, assuming that $\alpha_E=\alpha_M$,  $\gamma_E=\gamma_M$ we obtain that the E\"otv\"os parameter for the Earth and the Moon (\ref{e}) is equal to zero. So, the weak equivalence principle is preserved in noncommutative phase space of canonical type.

\section{Quantized space with preserved rotational and time-reversal symmetries and weak equivalence principle}\label{G_sec5}

\subsection{Rotationally-invariant noncommutative algebra of canonical type}
In six dimensional noncommutative phase space  of canonical type  (three dimensional configurational space and three dimensional momentum space) (\ref{33D1p})-(\ref{33D2p}) the rotational and time reversal symmetries are not preserved \cite{GnatenkoPRA}.

Algebra which is rotational invariant besides it is equivalent to noncommutative algebra of canonical type and  does not cause the time reversal symmetry breaking  was proposed in \cite{GnatenkoPRA}. It reads
\begin{eqnarray}
[X_{i},X_{j}]=i\hbar\theta_{ij}=ic_{\theta}\sum_k\varepsilon_{ijk}{p}^a_{k},\label{rotinv}\\{}
[X_{i},P_{j}]=i\hbar(\delta_{ij}+\gamma_{ij})=i\hbar\left(\delta_{ij}+\frac{c_{\theta}c_{\eta}}{4\hbar^2}({\bf {p}}^a\cdot{\bf {p}}^b)\delta_{ij}-\frac{c_{\theta}c_{\eta}}{4\hbar^2}p^{a}_j{p}^{b}_i\right),\label{for1001}\\{}
[P_{i},P_{j}]=i\hbar\eta_{ij}=i{c_{\eta}}\sum_k\varepsilon_{ijk}{p}^{b}_{k}.{}\label{rotinv1}
\end{eqnarray}
The algebra is constructed, considering tensors of noncommutativity defined as
\begin{eqnarray}
\theta_{ij}=\frac{c_{\theta}}{\hbar}\sum_k\varepsilon_{ijk}{p}^a_{k},\label{t1}\\
 \eta_{ij}=\frac{c_{\eta}}{\hbar}\sum_k\varepsilon_{ijk}{p}^b_{k},\label{t2}
  \end{eqnarray}
 here ${p}^{a}_i$, ${p}^{b}_i$ are additional momenta,  $c_{\theta}$, $c_{\eta}$ are constants, $\lim_{\hbar\rightarrow0}c_{\theta}/\hbar=\textrm{const}$, $\lim_{\hbar\rightarrow0}c_{\eta}/\hbar=\textrm{const}$ \cite{GnatenkoPRA}.
  From the symmetric representation of noncommutative coordinates and noncommutative momenta (see, for instance, \cite{Djemai,Bertolamih,Bertolami06} )
  follows  that parameters $\sigma_{ij}$ are defined as $\sigma_{ij}=\sum_k {\theta_{ik}\eta_{jk}}/{4}$.
So, using (\ref{t1}), (\ref{t2}), we obtain
 \begin{eqnarray}
 \sigma_{ij}=\frac{c_{\theta}c_{\eta}}{4\hbar^2}({\bf {p}}^a\cdot{\bf {p}}^b)\delta_{ij}-\frac{c_{\theta}c_{\eta}}{4\hbar^2}p^{a}_j{p}^{b}_i.\label{g1}
 \end{eqnarray}
 The symmetric representation for noncommutative coordinates and noncommutative momenta reads
  \begin{eqnarray}
X_{i}=x_{i}+\frac{1}{2}[{\bm \theta}\times{\bf p}]_i, \ \
P_{i}=p_{i}-\frac{1}{2}[{\bm \eta}\times {\bf x}]_i.\label{repp0}
\end{eqnarray}
Coordinates and momenta   $x_i$, $p_i$ satisfy the ordinary commutation relations
\begin{eqnarray}
 [x_i,x_j]=[p_i,p_j]=0, \ \
  [x_i,p_j]=i\hbar\delta_{ij}.
\end{eqnarray}
In (\ref{repp0})  we use notations ${\bm \theta}=(\theta_1, \theta_2, \theta_3)$, ${\bm \eta}=(\eta_1, \eta_2, \eta_3)$
\begin{eqnarray}
\theta_i=\sum_{jk}\frac{\varepsilon_{ijk}\theta_{jk}}{2}=\frac{c_{\theta}{p}_i^a}{\hbar}, \ \  \eta_i=\sum_{jk}\frac{\varepsilon_{ijk}\eta_{jk}}{2}=\frac{c_{\eta} p_i^b}{\hbar}.
\end{eqnarray}

 Additional momenta ${p}^a_i$, ${p}^b_i$ and additional coordinates ${a}_i$, ${b}_i$  satisfy the ordinary commutation relations
$[{a}_{i},{a}_{j}]=[{b}_{i},{b}_{j}]=[{a}_{i},{b}_{j}]=0$, $[{p}^{a}_{i},{p}^{a}_{j}]=[{p}^{b}_{i},{p}^{b}_{j}]=[{p}^{a}_{i},{p}^{b}_{j}]=0$,
$[{a}_{i},{p}^{a}_{j}]=[{b}_{i},{p}^{b}_{j}]=i\hbar\delta_{ij},$
$[{a}_{i},{p}^{b}_{j}]=[{b}_{i},{p}^{a}_{j}]=0,$
$[{a}_{i},X_{j}]=[{a}_{i},P_{j}]=[{p}^{b}_{i},X_{j}]=[{p}^{b}_{i},P_{j}]=0$. So, the tensors of noncommutativity commute with coordinates and momenta
\begin{eqnarray}
 [\theta_{ij},X_{k}]=[\theta_{ij},P_{k}]=[\eta_{ij},X_{k}]=[\eta_{ij},P_{k}]=0,{}\label{ro2} \\{}
[\sigma_{ij},X_{k}]=[\sigma_{ij},P_{k}]=0. \label{ro1}
\end{eqnarray}
The same relations (\ref{ro2}), (\ref{ro1}) are satisfied in the frame of noncommutative algebra of canonical type (\ref{33D1p})-(\ref{33D2p}).
In this sense algebra (\ref{rotinv})-(\ref{rotinv1}) is equivalent to (\ref{33D1p})-(\ref{33D2p})  \cite{GnatenkoPRA}.

To preserve the rotational symmetry additional coordinates and momenta $a_i$, $b_i$,  ${p}^a_i$, ${p}^b_i$ have to be governed by rotationally-symmetric systems. For reason of  simplicity in \cite{GnatenkoPRA} these systems were considered to be harmonic oscillators
\begin{eqnarray}
H^a_{osc}=\frac{({\bf p}^{a})^{2}}{2m_{osc}}+\frac{m_{osc}\omega^2_{osc}{\bf a}^{2}}{2}, \ \ H^b_{osc}=\frac{({\bf p}^{b})^{2}}{2m_{osc}}+\frac{m_{osc}\omega^2_{osc}{\bf b}^{2}}{2},\label{oscb}
\end{eqnarray}
with  $\sqrt{\hbar}/\sqrt{{m_{osc}\omega_{osc}}}=l_{P}$  and very large frequency $\omega_{osc}$  (oscillators put into the ground states remain in the states) \cite{GnatenkoPRA}.

 \subsection{Particle in gravitational field in noncommutative phase space with preserved rotational and time reversal symmetries}
Let us study the motion of a particle in uniform field in the frame of the algebra (\ref{rotinv})-(\ref{rotinv1}) and examine the weak equivalence principle. We consider the following Hamiltonian
\begin{eqnarray}
 H_p=\frac{{\bf P}^{2}}{2m}+mg X_1,
\end{eqnarray}
here $m$ is the mass of the particle, $g$  is the free fall acceleration. The $X_1$ axis is chosen to correspond to the field direction.
Coordinates and momenta of the particle satisfy relations of noncommutative algebra (\ref{rotinv})-(\ref{rotinv1}) which contain  additional momenta. So, to study the motion of the particle in gravitational field we have to take into account additional terms, corresponding to harmonic oscillators. Therefore, we consider the total Hamiltonian as follows
\begin{eqnarray}
 H=H_p+H^a_{osc}+H^b_{osc}.\label{total}
\end{eqnarray}
It is convenient to use representation   (\ref{repp0}) and rewrite the Hamiltonian in the following form
\begin{eqnarray}
 H=\frac{{\bf p}^{2}}{2m}+mg x_1-\frac{({\bm \eta}\cdot{\bf L})}{2m}+ \frac{mg}{2}[{\bm \theta}\times {\bf p}]_1+\nonumber\\+\frac{[{\bm \eta}\times{\bf x}]^2}{8m}+H^a_{osc}+H^b_{osc},\label{total}
\end{eqnarray}
here ${\bf L}=[{\bf x}\times{\bf p}]$.
The Hamiltonian also can be represented as
\begin{eqnarray}
H=H_0+\Delta H,\\
H_0=\langle H_p\rangle_{ab}+H^a_{osc}+H^b_{osc},\\
\Delta H= H-H_0=H_p-\langle H_p\rangle_{ab},
\end{eqnarray}
where $\langle...\rangle_{ab}=\langle\psi^{a}_{0,0,0}\psi^{b}_{0,0,0}|...|\psi^{a}_{0,0,0}\psi^{b}_{0,0,0}\rangle$,
$\psi^{a}_{0,0,0}$, $\psi^{b}_{0,0,0}$ are well known eigenfunctions of the harmonic oscillators  $H_{osc}^a$, $H_{osc}^b$ in the ground states.

For a particle in uniform field we have
\begin{eqnarray}
H_0=\frac{{\bf p}^{2}}{2m}+mg x_1+\frac{\langle\eta^2\rangle {\bf x}^2}{12m}+H^a_{osc}+H^b_{osc},\label{h0}\\
 \Delta H=-\frac{({\bm \eta}\cdot{\bf L})}{2m}+\frac{mg}{2}[{\bm \theta}\times {\bf p}]_1+\frac{[{\bm \eta}\times{\bf x}]^2}{8m}-\frac{\langle\eta^2\rangle {\bf x}^2}{12m}.\label{delta}
\end{eqnarray}
To find these expressions the following results are used
\begin{eqnarray}
\langle\psi^{a}_{0,0,0}|\theta_i|\psi^{a}_{0,0,0}\rangle=0, \ \
\langle\psi^{b}_{0,0,0}|\eta_i|\psi^{b}_{0,0,0}\rangle=0,\\
\langle\theta^2\rangle=\sum_i\langle\theta^2_i\rangle=\sum_i\frac{c_{\theta}^2}{\hbar^2}\langle\psi^{a}_{0,0,0}|({p}^{a}_i)^2|\psi^{a}_{0,0,0}\rangle=\frac{3c_{\theta}^2}{2l_P^2},\label{thetar2}\\
\langle\eta^2\rangle=\sum_i\langle\eta^2_i\rangle=\sum_i\frac{c_{\eta}^2}{\hbar^2}\langle\psi^{b}_{0,0,0}| ({p}^{b}_i)^2|\psi^{b}_{0,0,0}\rangle=\frac{3c_{\eta}^2}{2l_P^2}.
\end{eqnarray}
 In \cite{GnatenkoIJMPA18} it was shown that the corrections to $H_0$ caused by term $\Delta H$ vanish up to the second order of the perturbation theory.
 So, up to the second order in $\Delta H$ (or up to the second order in the parameters of noncommutativity) we can study Hamiltonian (\ref{h0}) and write the following equations of motion
\begin{eqnarray}
\dot x_i=\frac{p_i}{m},\ \
\dot p_i=-mg\delta_{i,1}-\frac{\langle\eta^2\rangle x_i}{6m}.\label{eqm1}
\end{eqnarray}
Solution of the equations with initial conditions $x_i(0)=x_{0i}$, $\dot x_i(t)=\upsilon_{0i}$ read \cite{GnatenkoEPL18}
\begin{eqnarray}\label{tr}
x_i(t)=\left(x_{0i}+6g{\frac{m^2}{\langle\eta^2\rangle}}\delta_{1,i}\right)\cos\left(\sqrt{\frac{\langle\eta^2\rangle}{6m^2}}t\right)+\nonumber\\+\upsilon_{0i}\sqrt{\frac{6m^2}{\langle\eta^2\rangle}}\sin\left(\sqrt{\frac{\langle\eta^2\rangle}{6m^2}}t\right)-6g{\frac{m^2}{\langle\eta^2\rangle}}\delta_{1,i}.
\end{eqnarray}
 From this result we can conclude that up to the second order in the parameters of noncommutativity the motion of a particle in uniform gravitational field is not affected by noncommutativity of coordinates. Also, it is worth noting that in limit $\langle\eta^2\rangle\rightarrow0$ from (\ref{tr}) we find well known result $x_i(t)=\delta_{1,i}gt^2/2+x_{0i}$ which corresponds to motion of a particle in gravitational field in the ordinary space.

It is important to mentioning that the trajectory of a particle in gravitational field  (\ref{tr}) depends on its mass. So, the weak equivalence principle is violated in noncommutative phase space of canonical type with preserved rotational and time reversal symmetries.

Note that if we consider the tensor of momentum noncommutativity to be dependent on mass as
  \begin{eqnarray}
   \eta_{ij}=\tilde{\alpha}m{\hbar}\sum_k\varepsilon_{ijk}{p}^b_{k}, \label{tc2}
 \end{eqnarray}
namely, if constant  $ c^{(n)}_{\eta}$ in  (\ref{t2}) satisfies  condition
  \begin{eqnarray}
\frac{c^{(n)}_{\eta}}{m_n}=\tilde{\alpha}=\textrm{const},\label{conde}
 \end{eqnarray}
(here $\tilde{\alpha}$ is the same for different particles), the motion of a particle in uniform field does not depend on mass and the weak equivalence principle is recovered \cite{GnatenkoEPL18,GnatenkoEPJP20}.

From  (\ref{conde}) follows that
\begin{eqnarray}\label{condit1}
\frac{\langle\eta^2\rangle}{m^2}=\frac{3\tilde{\alpha}^2}{2l_P^2}=B=\textrm{const},
\end{eqnarray}
and the trajectory of a particle reads
\begin{eqnarray}
x_i(t)=\left(x_{0i}+{\frac{6g}{B}}\delta_{1,i}\right)\cos\left(\sqrt{\frac{B}{6}}t\right)+\nonumber\\+\upsilon_{0i}\sqrt{\frac{6}{B}}\sin\left(\sqrt{\frac{B}{6}}t\right)-{\frac{6g}{B}}\delta_{1,i}.\label{tr1}
\end{eqnarray}

In the case of non-uniform gravitational field for a particle with mass $m$ we consider the following Hamiltonian
\begin{eqnarray}
H=H_p+H^a_{osc}+H^b_{osc},\  \
H_p=\frac{{P}^2}{2m}-\frac{G{\tilde M}m}{X},
 \end{eqnarray}
 here $ X=\sqrt{\sum_i X_i^2}$. The Hamiltonian $H_p$ written in representation (\ref{repp0}) up to the second order in the parameters of noncommutativity has the following form
\begin{eqnarray}
H_p=\frac{{p}^2}{2m}-\frac{G{\tilde M}m}{x}-\frac{({\bm \eta}\cdot{\bf L})}{2m}+\frac{[{\bm \eta}\times{\bf x}]^2}{8m}-\nonumber\\-\frac{G{\tilde M}m}{\sqrt{x^2-({\bm \theta}\cdot{\bf L})+\frac{[{\bm \theta}\times{\bf p}]^2}{4}}}=
\frac{{p}^2}{2m}-\frac{G{\tilde M}m}{x}-\frac{({\bm \eta}\cdot{\bf L})}{2m}+\frac{[{\bm \eta}\times{\bf x}]^2}{8m}-\nonumber\\-\frac{G{\tilde M}m}{2x^3}({\bm \theta}\cdot{\bf L})-\frac{3G{\tilde  M}m}{8x^5}({\bm \theta}\cdot{\bf L})^2+\frac{G{\tilde M}m}{16} \left(\frac{1}{x^2}[{\bm \theta}\times{\bf p}]^2\frac{1}{x}+\frac{1}{x}[{\bm \theta}\times{\bf p}]^2\frac{1}{x^2}+\frac{\hbar^2}{x^7}[{\bm \theta}\times{\bf x}]^2\right),\nonumber\\\label{hs}
 \end{eqnarray}
where $x=|{\bf x}|$ (the details of calculations of the expansion can be found in \cite{GnatenkoPLA14}).
So, for $\Delta H$ we have
 \begin{eqnarray}
\Delta H=-\frac{({\bm \eta}\cdot{\bf L})}{2m}+\frac{[{\bm \eta}\times{\bf x}]^2}{8m}-\frac{\langle\eta^2\rangle { x}^2}{12m}-\frac{G{\tilde  M} m}{2x^3}({\bm \theta}\cdot{\bf L})+\frac{G{\tilde  M}m L^2\langle\theta^2\rangle}{8 x^5}+\nonumber\\+\frac{G{\tilde  M} m}{16}\left(\frac{1}{x^2}[{\bm \theta}\times{\bf p}]^2\frac{1}{x}+\frac{1}{x}[{\bm \theta}\times{\bf p}]^2\frac{1}{x^2}+\frac{\hbar^2}{x^7}[{{\bm \theta}}\times{\bf x}]^2\right)-\frac{3G{\tilde M}m}{8x^5}({\bm \theta}\cdot{\bf L})^2-\nonumber\\-\frac{G{\tilde  M}m\langle\theta^2\rangle}{24}\left(\frac{1}{x^2} p^2\frac{1}{x}+\frac{1}{x}p^2\frac{1}{x^2}+\frac{\hbar^2}{x^5}\right).\label{dd}
 \end{eqnarray}
Up to the second order in the parameters of noncommutativity to study the motion of a particle in nonuniform gravitational field we  can consider the following Hamiltonian
\begin{eqnarray}
H_0=\frac{{ p}^2}{2m}-\frac{G{\tilde  M}m}{x}+\frac{\langle\eta^2\rangle  x^2}{12m}-\frac{G{\tilde  M}m L^2\langle\theta^2\rangle}{8 x^5}+\nonumber\\+\frac{G{\tilde  M}m\langle\theta^2\rangle}{24}\left(\frac{2}{x^3} p^2+\frac{6i\hbar}{x^5}({\bf x}\cdot{\bf p}) -\frac{\hbar^2}{x^5}\right)+H^a_{osc}+H^b_{osc},\label{he}
\end{eqnarray}
 and find the following equations of motion
\begin{eqnarray}
\dot{{\bf x}}=\frac{{\bf p}}{m}-\frac{G{\tilde  M}m\langle\theta^2\rangle}{12}\left(\frac{1}{x^3}{\bf p}-\frac{3{\bf x}}{x^5}({\bf x}\cdot{\bf p})\right),\label{r}\\
\dot{{\bf p}}=-\frac{G{\tilde  M}m {\bf x}}{x^3}-\frac{\langle\eta^2\rangle  {\bf x}}{6m}-\frac{G{\tilde M}m\langle\theta^2\rangle}{4}\left(\frac{1}{x^5}({\bf x}\cdot{\bf p}){\bf p}-\frac{2{\bf x}}{x^5}p^2+\frac{5{\bf x}}{2x^7}L^2+\frac{5\hbar^2{\bf x}}{6x^7}-\frac{5i\hbar}{x^7}{\bf x}({\bf x}\cdot{\bf p})\right).\nonumber\\\label{p}
\end{eqnarray}
These equations in the classical limit  ($\hbar\rightarrow0$) transform to
\begin{eqnarray}
\dot{{\bf x}}={\bm p}^{\prime}-\frac{G{\tilde  M}m^2\langle\theta^2\rangle}{12}\left(\frac{1}{x^3}{\bm p}^{\prime}-\frac{3{\bf x}}{x^5}({\bf x}\cdot{\bm p}^{\prime}) \right),\label{rv}\\
\dot{{\bm p}}^{\prime}=-\frac{G {\tilde M} {\bf x}}{x^3}-\frac{\langle\eta^2\rangle {\bf x}}{6m^2}-\frac{G {\tilde M} m^2\langle\theta^2\rangle}{4}\left(\frac{1}{x^5}({\bf x}\cdot{\bm p}^{\prime}){\bm p}^{\prime}-\frac{2{\bf x}}{x^5}(p^{\prime})^2+\frac{5{\bf x}}{2x^7}[{\bf x}\times{\bm p}^{\prime}]^2\right),\label{pv}
\end{eqnarray}
here ${\bm p}^{\prime}={{\bf p}}/{m}$ \cite{GnatenkoEPL18}.
Let us consider dependence of the tensor of coordinates noncommutativity on mass as follows

\begin{eqnarray}
c^{(n)}_{\theta}m_n=\tilde{\gamma}=\textrm{const},\label{condt}\\
\theta_{ij}=\frac{\tilde{\gamma}}{m}\hbar\sum_k\varepsilon_{ijk}p^{a}_{k},\\
\langle\theta^2\rangle m^2=\frac{3\tilde{\gamma}^2}{2l_P^2}=A=\textrm{const},
\end{eqnarray}
where constants  $A$, $\tilde{\gamma}$ are the same for different particles.
So, in the case when the relations  (\ref{conde}), (\ref{condt}) hold, form
 (\ref{r}), (\ref{p}) we obtain
\begin{eqnarray}
\dot{{\bf x}}={\bm p}^{\prime}-\frac{G {\tilde  M} B}{12}\left(\frac{1}{x^3}{\bm p}^{\prime}-\frac{3{\bf x}}{x^5}({\bf x}\cdot{\bm p}^{\prime}) \right),\label{rvx}\\
\dot{{\bm p}}^{\prime}=-\frac{G {\tilde  M} {\bf x}}{x^3}-\frac{B{\bf x}}{6}-\frac{G {\tilde M} A}{4}\left(\frac{1}{x^5}({\bf x}\cdot{\bm p}^{\prime}){\bm p}^{\prime}-\frac{2{\bf x}}{x^5}(p^{\prime})^2+\right.\nonumber\\\left.+\frac{5{\bf x}}{2x^7}[{\bf x}\times{\bm p}^{\prime}]^2+\frac{5\hbar^2{\bf x}}{6m^2x^7}-\frac{5i\hbar}{mx^7}{\bf x}({\bf x}\cdot{\bm p}^{\prime})\right).\label{pvx}
\end{eqnarray}
In the classical limit on the basis of (\ref{rvx}), (\ref{pvx}) we find
\begin{eqnarray}
\dot{{\bf x}}={\bm p}^{\prime}-\frac{G{\tilde  M}A}{12}\left(\frac{1}{x^3}{\bm p}^{\prime}-\frac{3{\bf x}}{x^5}({\bf x}\cdot{\bm p}^{\prime}) \right),\label{ra}\\
\dot{{\bm p}}^{\prime}=-\frac{G {\tilde M} {\bf x}}{x^3}-\frac{B{\bf x}}{6}-\nonumber\\-\frac{G {\tilde M} A}{4}\left(\frac{1}{x^5}({\bf x}\cdot{\bm p}^{\prime}){\bm p}^{\prime}-\frac{2{\bf x}}{x^5}(p^{\prime})^2+\frac{5{\bf x}}{2x^7}[{\bf x}\times{\bm p}^{\prime}]^2\right).\label{pa}
\end{eqnarray}
 The equations of motion of a particle in gravitational field in quantum case (\ref{rvx}), (\ref{pvx}) depend on the ratio ${\hbar}/{m}$, as it has to be. This is caused by dependence of commutation relation on mass $[{\bf x}, {\bm p}^{\prime}]=i\hbar\hat{I}/m$ \cite{Greenberger68}.
 Classical equations of motion  (\ref{ra}), (\ref{pa}) do not depend on mass. So, the weak equivalence principle is satisfied in noncommutative phase space with preserved rotational and time reversal symmetries if tensors of noncommutativity are related with mass (\ref{conde}), (\ref{condt}) \cite{GnatenkoEPL18}.

It is worth noting that the considered in this section conditions (\ref{conde}), (\ref{condt}) are in agreement to that presented in  section \ref{G_sec3} (\ref{ondN2}), (\ref{ondN}) to recover  the weak equivalence principle in noncommutative phase space of canonical type.

\section{Weak equivalence principle in the frame of noncommutative algebra of Lie type}\label{G_sec6}

 \subsection{Lie algebra with space coordinates commuting to time and the weak equivalence principle}
Let us study the motion of a particle in gravitational field in a space with noncommutativity of Lie type in the case when space coordinates commute to time
 \begin{eqnarray}
 [X_i,X_j]=\frac{i\hbar t}{\kappa}\left(\delta_{i \rho}\delta_{j\tau}-\delta_{i\tau}\delta_{j\rho}\right),{}\label{nt}\\{}
  [X_i,P_j]=i\hbar\delta_{ij},{}\ \
  [P_i,P_j]=0,\label{nt1}
 \end{eqnarray}
 here $i,j=(1,2,3)$, indexes $\rho$, $\tau$ are fixed and different, $\kappa$  is a parameter  \cite{DaszkiewiczPRD,DaszkiewiczMPLA08}.
 The deformed Poisson brackets corresponding to  (\ref{nt})-(\ref{nt1})  are the following
 \begin{eqnarray}
 \{X_i,X_j\}=\frac{t}{\kappa}\left(\delta_{i \rho}\delta_{j\tau}-\delta_{i\tau}\delta_{j\rho}\right),{}\label{tn}\\{}
  \{X_i,P_j\}=\delta_{ij},{}\ \
  \{P_i,P_j\}=0,\label{tn1}
 \end{eqnarray}
(see \cite{DaszkiewiczPRD}).

For a particle with mass $m$ in gravitational field $V=V(X_1,X_2,X_3)$  the Hamiltonian reads
 \begin{eqnarray}
H=\frac{{\bf P}^2}{2m}+mV(X_1,X_2,X_3).\label{h}
\end{eqnarray}
Taking into account (\ref{tn}), (\ref{tn1}), we can write equations of motion as follows
 \begin{eqnarray}
\dot{X}_i=\{X_i,H\}=\frac{P_i}{m}+\frac{tm}{\kappa}\frac{\partial V}{\partial X_k}\left(\delta_{i\rho}\delta_{k\tau}-\delta_{i\tau}\delta_{k\rho}\right),\label{xcd}\\
\dot{P}_i=\{P_i,H\}=-m\frac{\partial V}{\partial X_i},
 \end{eqnarray}
 (see  \cite{DaszkiewiczPRD,GnatenkoPRD}).
  Note that  in (\ref{xcd}) because of noncommutativity of Lie type we have term proportional to mass $m$. Therefore the weak equivalence principle is violated.
Similarly as in noncommutative space of canonical type, let us consider dependence of the parameter on noncommutative algebra on mass and write the following condition
\begin{eqnarray}
\frac{\kappa}{m}=\gamma_{\kappa}=\textrm{const},\label{cond}
\end{eqnarray}
here $\gamma_{\kappa}$ does not depend on mass and is the same for different particles.
Taking into account relation (\ref{cond}), the equations of motion of a particle in gravitational field can be rewritten as
 \begin{eqnarray}
\dot{X}_i=P^{\prime}_i+\frac{t}{\gamma_{\kappa}}\frac{\partial V}{\partial X_k}\left(\delta_{i\rho}\delta_{k\tau}-\delta_{i\tau}\delta_{k\rho}\right),\ \
\dot{P}^{\prime}_i=-\frac{\partial V}{\partial X_i},\label{xdc1}
 \end{eqnarray}
where $P^{\prime}_i=P_i/m$. So, on the basis of the obtained result we have that $X_i(t)$, $P^{\prime}_i(t)$ do not depend on mass and the weak equivalence principle is recovered if condition (\ref{cond}) is satisfied \cite{GnatenkoPRD}.

Let us also study the case of motion of a composite system in gravitational field and examine the weak equivalence principle.
For coordinates and momenta of different particles the noncommutative algebra of Lie type (\ref{tn}), (\ref{tn1})  can be generalized as
\begin{eqnarray}
\{X_{i}^{(a)},X_{j}^{(b)}\}=\frac{ t}{\kappa_a}\left(\delta_{i \rho}\delta_{j\tau}-\delta_{i\tau}\delta_{j\rho}\right)\delta_{ab},\label{al0}\\{}
\{X_{i}^{(a)},P_{j}^{(b)}\}=\delta_{ab}\delta_{ij},\ \
\{P_{i}^{(a)},P_{j}^{(b)}\}=0,\label{al1}
\end{eqnarray}
here $X_{i}^{(a)}$, $P_{i}^{(a)}$, ${\kappa_a}$ are coordinates, momenta and parameters of noncommutative algebra corresponding to the particle with index $a$ \cite{GnatenkoPRD}.
Noncommutative algebra for coordinates and momenta of the center-of-mass, coordinates and momenta of the relative motion introduced in the  traditional way  ($\tilde{{\bf P}}=\sum_{a}{\bf P}^{(a)}$, $\tilde{{\bf X}}=\sum_{a}\mu_{a}{\bf X}^{(a)},$
$\Delta{\bf P}^{{a}}={\bf P}^{(a)}-\mu_{a}\tilde{{\bf P}}$, ${\Delta\bf X}^{(a)}={\bf X}^{(a)}-\tilde{{\bf X}}$, $\mu_a=m_{a}/M$, $M=\sum_{a}m_a$) is the following
\begin{eqnarray}
\{\tilde{X}_i,\tilde{X}_j\}= t\sum_{a}\frac{\mu_{a}^{2}}{\kappa_a}\left(\delta_{i \rho}\delta_{j\tau}-\delta_{i\tau}\delta_{j\rho}\right),\label{07}\\{}
\{\tilde{X}_i,\tilde{P}_j\}=\delta_{ij}, \ \ \{\tilde{P}_i,\tilde{P}_j\}=0 \label{08}\label{007}\\{}
\{\Delta{X}_i^{(a)},\Delta{X}_j^{(b)}\}= t\left(\frac{\delta^{ab}}{\kappa_a}-\frac{\mu_{a}}{\kappa_{a}}-\frac{\mu_{b}}{\kappa_{b}}+\sum_{c}\frac{\mu_{c}^{2}}{\kappa_c}\right)\left(\delta_{i \rho}\delta_{j\tau}-\delta_{i\tau}\delta_{j\rho}\right),\\
\{\Delta{X}^{(a)}_i,\Delta{P}^{(b)}_j\}=\delta_{ab}-\mu_b,\\{}
\{\Delta{X}_i^{(a)},\tilde{X}_j\}= t \left(\frac{\mu_{a}}{\kappa_{a}}-\sum_{c}\frac{\mu_{c}^{2}}{\kappa_c}\right)\left(\delta_{i \rho}\delta_{j\tau}-\delta_{i\tau}\delta_{j\rho}\right),\label{cr}\\{}
\{\Delta{P}_i^{(a)},\Delta{P}_i^{(b)}\}=\{\tilde{P}_i,\Delta{P}_j^{(b)}\}=0.{}\label{09}
\end{eqnarray}
The Poisson brackets for coordinates of the center-of-mass and coordinates of the relative motion vanish
\begin{eqnarray}
\{\Delta{X}_i^{(a)},\tilde{X}_j\}=0,\label{van}
\end{eqnarray}
if the parameters of noncommutative algebra are determined by mass as (\ref{cond}) \cite{GnatenkoPRD}. Namely, if relation $\kappa_a=m_a\gamma_{\kappa}$ is satisfied. Also, in this case the effective parameter of noncommutativity depends on the total mass of the system and is independent of its composition
\begin{eqnarray}
\tilde{\theta}^0_{ij}=\sum_{a}\frac{\mu_{a}^{2}}{\kappa_a}\left(\delta_{i \rho}\delta_{j\tau}-\delta_{j\tau}\delta_{i\rho}\right)=\frac{1}{{\kappa}_{eff}}\left(\delta_{i \rho}\delta_{j\tau}-\delta_{i\tau}\delta_{j\rho}\right)=\nonumber\\=\frac{1}{\gamma_{\kappa}M}\left(\delta_{i \rho}\delta_{j\tau}-\delta_{j\tau}\delta_{i\rho}\right)\label{eff2}.
\end{eqnarray}

Let us study the motion of a composite system of mass $M$ in gravitational field in the space with Lie algebraic noncommutativity (\ref{tn}), (\ref{tn1}) on the basis of the obtained results. The Hamiltonian reads
\begin{eqnarray}
H=\frac{{\bf \tilde{P}}^2}{2M}+MV(\tilde{X}_1,\tilde{X}_2,\tilde{X}_3)+H_{rel}.\label{body}
\end{eqnarray}
The term  $H_{rel}$ corresponds to the relative motion, $\tilde{X}_i$, $\tilde{P}_i$  are coordinates and momenta of the center-of-mass of the composite system that are defined in the traditional way.

Considering condition on the parameter of noncommutative algebra (\ref{cond}), we have (\ref{van}) and
\begin{eqnarray}
\left\{\frac{{\bf \tilde{P}}^2}{2M}+MV(\tilde{X}_1,\tilde{X}_2,\tilde{X}_3),H_{rel}\right\}=0,
\end{eqnarray}
So, for a composite system we can write the following equations of motion in the gravitational field
\begin{eqnarray}
\dot{{\tilde X}}_i=\frac{\tilde{P}_i}{M}+tM\sum_{a}\frac{\mu_{a}^{2}}{\kappa_a}\left(\delta_{i \rho}\delta_{j\tau}-\delta_{i\tau}\delta_{j\rho}\right)\frac{\partial V}{\partial \tilde{X}_j}=
\tilde{P}^{\prime}_i+t\sum_{a}\frac{1}{\gamma_\kappa}\left(\delta_{i \rho}\delta_{j\tau}-\delta_{i\tau}\delta_{j\rho}\right)\frac{\partial V}{\partial \tilde{X}_j},\label{zxd5}\\
\dot{{\tilde P}}_i=-M\frac{\partial V}{\partial {\tilde X}_i}=-\frac{\partial V}{\partial {\tilde X}_i},\label{zxd6}
 \end{eqnarray}
here $\tilde{P}^{\prime}_i=\tilde{P}_i/M$. From equations  (\ref{zxd5}), (\ref{zxd6}) follows that expressions for ${\tilde X}_i(t)$, $\tilde{P}^{\prime}_i(t)$ do not depend on the mass of the composite system and its composition. So, the weak equivalence principle is recovered in the frame of the algebra  (\ref{tn}), (\ref{tn1})  due to condition (\ref{cond}) \cite{GnatenkoPRD}.

 \subsection{Preserving of the weak equivalence principle in the general case of Lie algebraic noncommutativity}

In more general case of noncommutative algebra of Lie type the Poisson brackets are the following
\begin{eqnarray}
 \{X_i,X_j\}=\theta^0_{ij}t+\theta^k_{ij}X_k,{}\label{gen}\\{}
  \{X_i,P_j\}=\delta_{ij}+\bar{\theta}^k_{ij}X_k+\tilde{\theta}^k_{ij}P_k,{}\ \
  \{P_i,P_j\}=0,\label{gen1}
 \end{eqnarray}
  here $i,j,k=(1,2,3)$,  $\theta^0_{ij}$, $\theta^k_{ij}$, $\bar{\theta}^k_{ij}$, $\tilde{\theta}^k_{ij}$ are constants, $\theta^0_{ij}=-\theta^0_{ji}$, $\bar{\theta}^k_{ij}=-\bar{\theta}^k_{ji}$, $\tilde{\theta}^k_{ij}=-\tilde{\theta}^k_{ij}$ \cite{Miao}.
 These constants have to be chosen to satisfy the Jacobi identity.  This issue was studied in \cite{Miao}. The author of the paper considered the following algebras of Lie type
\begin{eqnarray}
\{X_k,X_{\gamma}\}=-\frac{t}{\kappa}+\frac{X_l}{\tilde{\kappa}},\ \  \{X_l,X_{\gamma}\}=\frac{t}{\kappa}-\frac{X_k}{\tilde{\kappa}}, \label{gsts} {}\\ {}
\{X_k,X_{l}\}=\frac{t}{\kappa}, \ \ \{P_k,X_{\gamma}\}=\frac{P_l}{\tilde{\kappa}},{}\\{}
 \{P_l,X_{\gamma}\}=-\frac{P_k}{\tilde{\kappa}},  \ \  \{X_i,P_j\}=\delta_{ij}, {}\\{}
    \{X_{\gamma},P_{\gamma}\}=1  \ \ \{P_m,P_n\}=0,{} \label{gsts1}
\end{eqnarray}
and the second ones
\begin{eqnarray}
\{X_k,X_{\gamma}\}=-\frac{t}{\kappa}+\frac{X_l}{\tilde{\kappa}}, \ \  \{X_l,X_{\gamma}\}=\frac{t}{\kappa}-\frac{X_k}{\tilde{\kappa}}, \label{gsts2} {}\\ {}
\{X_k,X_{l}\}=0, \ \ \{P_k,X_{\gamma}\}=\frac{X_l}{\bar{\kappa}}+\frac{P_l}{\tilde{\kappa}},{}\\ {} \{P_l,X_{\gamma}\}=\frac{X_k}{\bar{\kappa}}-\frac{P_k}{\tilde{\kappa}},  \ \ \{X_i,P_j\}=\delta_{ij}, {}\\{}
   \{X_{\gamma},P_{\gamma}\}=1,{}\ \ \{P_m,P_n\}=0,{} \label{gsts12}
\end{eqnarray}
The algebras correspond to the cases when parameters of noncommutativity satisfy the following relations
  \begin{eqnarray}
  \theta^0_{kl}=-\theta^0_{k\gamma}=\frac{1}{\kappa}, \ \  \theta^0_{l\gamma}=\frac{1}{\kappa},  \\ \theta^l_{k\gamma}=-\theta^k_{l\gamma}=\tilde{\theta}^l_{k\gamma}=-\tilde{\theta}^k_{l\gamma}=\frac{1}{\tilde{\kappa}},
  \end{eqnarray}
and
\begin{eqnarray}
\theta^0_{l\gamma}=-\theta^0_{k\gamma}=\frac{1}{\kappa}, \ \   \theta^l_{k\gamma}=-\theta^k_{l\gamma}=\frac{1}{\tilde{\kappa}}, \\  \tilde{\theta}^l_{k\gamma}=-\tilde{\theta}^k_{l\gamma}=\frac{1}{\tilde{\kappa}}, \\ \bar{\theta}^l_{k\gamma}=-\bar{\theta}^k_{l\gamma}=\frac{1}{\bar{\kappa}},
\end{eqnarray}
respectively.

 For a particle in gravitational field (\ref{h}) taking into account  (\ref{gen}), (\ref{gen1}) we obtain that the equations of motion depend on mass
\begin{eqnarray}
\dot{X}_i=\frac{P_i}{m}+\bar{\theta}^k_{ij}\frac{P_j X_k}{m}+\tilde{\theta}^k_{ij}\frac{P_j P_k}{m}+m(\theta^{0}_{ij}t+\theta^{k}_{ij}X_k)\frac{\partial V}{\partial X_j},\label{xd}\\
\dot{P}_i=-m\frac{\partial V}{\partial X_i}-m(\bar{\theta}^{k}_{ij}X_{k}+\tilde{\theta}^{k}_{ij}P_k)\frac{\partial V}{\partial X_j}.
 \end{eqnarray}
Due to  relations of the parameters of noncommutativity with mass proposed in \cite{GnatenkoPRD}
 \begin{eqnarray}
\theta^{0(a)}_{ij}m_a={\gamma^{0}_{ij}}=\textrm{const},\ \ \theta^{k(a)}_{ij}m_a=\gamma^{k}_{ij}=\textrm{const},\label{cond2}\\
\tilde{\theta}^{k(a)}_{ij}m_a={\tilde{\gamma}^{k}_{ij}}=\textrm{const}, \ \ \bar{\theta}^{k(a)}_{ij}=\bar{\theta}^{k}_{ij}. \label{cond3}
\end{eqnarray}
we obtain
\begin{eqnarray}
\dot{X}_i={P^{\prime}_i}+\bar{\theta}^k_{ij}{P^{\prime}_j X_k}+\tilde{\gamma}^k_{ij}{P^{\prime}_j P^{\prime}_k}+(\gamma^{0}_{ij}t+\gamma^{k}_{ij}X_k)\frac{\partial V}{\partial X_j},\label{eq1}\\
\dot{P}^{\prime}_i=-\frac{\partial V}{\partial X_i}-(\bar{\theta}^{k}_{ij}X_{k}+\tilde{\gamma}^{k}_{ij}P^{\prime}_k)\frac{\partial V}{\partial X_j}.\label{eq2}
 \end{eqnarray}
 Here constants $\gamma^{0}_{ij}$, $\gamma^{k}_{ij}$, $\tilde{\gamma}^{k}_{ij}$  do not depend on mass $\gamma^{0}_{ij}=-\gamma^{0}_{ji}$, $\gamma^{k}_{ij}=-\gamma^{k}_{ji}$, $\tilde{\gamma}^{k}_{ij}=-\tilde{\gamma}^{k}_{ji}$, $P^{\prime}_i=P_i/m$.
 So, if conditions (\ref{cond2}), (\ref{cond3}) hold the weak equivalence principle is preserved in noncommutative space of Lie type (\ref{gen}), (\ref{gen1}).

Let us also study motion of a composite system in gravitational field in the space (\ref{gen}), (\ref{gen1}) and examine the weak equivalence principle. The noncommutative algebra (\ref{gen}), (\ref{gen1})  can be generalized for  coordinates and momenta of different particles $X^{(a)}_i$, $P^{(a)}_i$ (index $a$ label a particle) as
\begin{eqnarray}
\{X^{(a)}_i,X^{(b)}_j\}=\delta_{ab}\theta^{0(a)}_{ij}t+\delta_{ab}\theta^{k(a)}_{ij}X^{(a)}_k,{} \label{c01}\\{}
\{X^{(a)}_i,P^{(b)}_j\}=\delta_{ab}\delta_{ij}+\delta_{ab}\bar{\theta}^{k(a)}_{ij}X^{(a)}_k+\delta_{ab}\tilde{\theta}^{k(a)}_{ij}P^{a}_k,{}\\{}
\{P^{(a)}_i,P^{(b)}_j\}=0,\label{c001}
 \end{eqnarray}
 $\theta^{0(a)}_{ij}$, $\theta^{k(a)}_{ij}$, $\bar{\theta}^{k(a)}_{ij}$, $\tilde{\theta}^{k(a)}_{ij}$ are parameters of noncommutative algebra corresponding to a particle with index $a$ \cite{GnatenkoPRD}. The relations of noncommutative algebra for coordinates and momenta of the center-of-mass read
\begin{eqnarray}
\{\tilde{X}_i,\tilde{X}_j\}=\sum_a\mu_a^2\theta^{0(a)}_{ij}t+\sum_a\mu_a^2\theta^{k(a)}_{ij}X^{(a)}_k,{} \label{c11}\\{}
\{\tilde{X}_i,\tilde{P}_j\}=\delta_{ij}+\sum_{a}\mu_a\bar{\theta}^{k(a)}_{ij}X^{(a)}_k+\sum_a\mu_a\tilde{\theta}^{k(a)}_{ij}P^{a}_k,{}\label{c111}\\{}
\{\tilde{P}_i,\tilde{P}_j\}=0.\label{c1}
 \end{eqnarray}
 Note that the  relations (\ref{c11}), (\ref{c1}) do not reproduce relations of Lie algebra (\ref{gen})-(\ref{gen1}).  In the right-hand side of (\ref{c11}), (\ref{c111}) we do not have coordinates and momenta of the center-of-mass. It is important to mention that the problem is solved due to conditions (\ref{cond2}), (\ref{cond3}) \cite{GnatenkoPRD}. For coordinates and momenta of the center-of-mass one obtains relations of noncommutative algebra of Lie type
\begin{eqnarray}
\{\tilde{X}_i,\tilde{X}_j\}=\theta^{0(eff)}_{ij}t+\theta^{k(eff)}_{ij}\tilde{X}_k,{} \label{cm111}\\{}
\{\tilde{X}_i,\tilde{P}_j\}=\delta_{ij}+\bar{\theta}^{k}_{ij}\tilde{X}_k+\tilde{\theta}^{k(eff)}_{ij}\tilde{P}_k,{}\label{cm12}
 \end{eqnarray}
 with parameters
 \begin{eqnarray}
\theta^{0(eff)}_{ij}=\frac{\gamma^{0}_{ij}}{M}, \ \ \theta^{k(eff)}_{ij}=\frac{\gamma^{k}_{ij}}{M}, \ \ \tilde\theta^{k(eff)}_{ij}=\frac{\tilde{\gamma}^{k}_{ij}}{M},
 \end{eqnarray}
here $M=\sum_am_a$ is the total mass of the system \cite{GnatenkoPRD}.

So, on the basis of these results one can write the equations of motion of composite system in gravitational field  in quantized space with algebra (\ref{gen}), (\ref{gen1}).
Introducing notation $\tilde{P}^{\prime}_i=\tilde{P}_i/M$ for a composite system in gravitational field we find
\begin{eqnarray}
\dot{{\tilde X}}_i={\tilde{P}^{\prime}_i}+\left(\bar{\theta}^{k}_{ij}\tilde{X}_k+\tilde{\gamma}^{k}_{ij}\tilde{P}^{\prime}_k\right) \tilde{P}^{\prime}_j+\left(\gamma^{0}_{ij}t+\gamma^{k}_{ij}\tilde{X}_k\right)\frac{\partial V}{\partial \tilde{X}_j},\label{eq111}\\
\dot{{\tilde P}}^{\prime}_i=-\frac{\partial V}{\partial {\tilde X}_i}-\left(\bar{\theta}^{k}_{ij}\tilde{X}_k+\tilde{\gamma}^{k}_{ij}\tilde{P}^{\prime}_k\right)\frac{\partial V}{\partial\tilde{X}_j}.\label{eq211}
\end{eqnarray}
Writing (\ref{eq111}), (\ref{eq211}) we assume that influence of the relative motion on  the motion of the center-of-mass of the system can be neglected.
 Equations of motion of a composite system in gravitational field (\ref{eq111}), (\ref{eq211}) do not depend on its total mass, masses of particles forming it, its composition. So, the weak equivalence principle is  preserved in a general case of noncommutative algebra of Lie type (\ref{gen}), (\ref{gen1}) due to relations (\ref{cond2}), (\ref{cond3}) \cite{GnatenkoPRD}.

\section{Conclusions}\label{G_sec7}

We have examined quantum spaces with different deformed Heisenberg algebras (noncommutative algebra of canonical type, noncommutative algebra of Lie type, Snyder algebra, Kempf algebra and their generalizations). A motion of a particle in gravitational field has been studied in the frame of the deformed algebras and the implementation of the weak equivalence principle has been analyzed.

We have concluded that different types of deformation of the commutation relations for coordinates and momenta (canonical, Lie and nonlinear deformations) lead to dependence of motion of a particle (composite system) in gravitational filed on mass and its composition. Therefore the weak equivalence principle is violated. The principle is violated even in the case when the gravitational mass is equal to the inertial mass. It is worth stressing that the deformation of the algebra leads to grate violation of the principle that can be easily seen at an experiment. But from the observations we know that the weak equivalence principle is preserved with hight accuracy.
The problem is solved if one considers parameters of deformed algebras to be dependent on mass. In this case the motion of a particle (composite system) in gravitational field does not depend on its mass and composition and the weak equivalence principle is recovered.

It is important to add that the idea to relate parameters of deformed algebra to  mass is also important for preserving of the properties of the kinetic energy (additivity property, independence on composition) in quantum space and therefore for recovering of the law of conservation of energy. Also, in the case when parameters of deformed algebra depend on mass the problem of description of motion of a composite system in a space with minimal length is solved. The problem is well known in the literature as the soccer-ball problem.

So, the idea of dependence of parameters of deformed algebras on mass leads to solving of fundamental problems in a space with minimal length, among them violation of the weak equivalence principle, nonadditivity of the kinetic energy and its dependence on composition, the soccer-ball problem.

\section*{Acknowledgments}
This work was partly supported by the Projects $\Phi\Phi-27\Phi$ (0122U001558), $\Phi\Phi$-11Hp (No. 0121U100058) from the Ministry of Education and Science of Ukraine.


\newpage
\begin{center}
\begin{large}{\bf Деформовані алгебри Гайзенберга різних типів зі збереженим принципом еквівалентності}
\end{large}
\end{center}

\centerline { Х. П. Гнатенко, В. М. Ткачук}
\medskip

\centerline {\small  \it Кафедра теоретичної фізики імені професора Івана Вакарчука,}
\centerline {\small \it Львівський національний університет імені Івана Франка,}
\centerline {\small \it Вул. Драгоманова 12, 79005 Львів, Україна}

\medskip
\centerline {{\bf Анотація}}
Розглядається ідея опису квантованості простору (існування кванта довжини) за допомогою модифікації комутаціних співвідношень для операторів координат та операторів імпульсів. Вивчаються різні типи деформації алгебри Гайзенберга, а саме канонічна (комутатори координат та імпульсів дорівнюють константам), типу Лі (комутатори координат та імпульсів пропорційні до координат та імпульсів) та нелінійна деформації (комутатори координат та імпульсів дорівнюють нелінійній функції цих координат та імпульсів). Досліджуються некомутативна алгебра з некомутативністю координат та некомутативністю імпульсів канонічного типу, некомутативна алгебра Лі типу, алгебра Снайдера, алгебра Кемпфа та їх узагальнення на випадок, коли комутатор координат та імпульсів дорівнює довільній функцій, яка залежить від імпульсів. У рамках різних деформованих алгебр вивчається рух частинки (макроскопічного тіла) у гравітаційному полі та аналізується виконання слабкого принципу еквівалентності. Показано, що у квантованому просторі рух у гравітаціному полі залежить від маси та композиції. Параметр Етвеша не дошрівнює нулеві навіть у випадку, коли інерційна маса дорівнює гравітаційній. Слабкий принцип еквівалентності порушується у квантованому просторі, при чому деформація комутаційних співвідношень для операторів координат та операторів імпульсів зумовлює значні поправки до параметра Етвеша, які легко можуть спостерігатися на експерименті. З іншого боку, відповідно до експериментальних даних слабкий принцип еквівалентності виконується з великою точністю. Цю проблему можна розв'язати, припустивши, що параметри деформованих алгебр залежать від маси. Така ідея дозволяє відновити слабкий принцип еквівалентності, а також зберегти властивості кінетичної енергії, розв'язати проблему опису руху макроскопічного тіла (ця проблема добре відома у літературі під назвою проблема футбольного м'яча) у квантованому просторі. Отже, залежність параметрів деформації від маси є важливою для побудови теорії квантованого простору зі збереженими фундаментальними законами та принципами.
\end{document}